\documentclass[12pt,3p]{elsarticle}
\usepackage{setspace}
\usepackage{amssymb}
\usepackage{gensymb}
\usepackage{rotating}
\usepackage{tabularx} 
\usepackage{float}
\usepackage{booktabs}
\usepackage{url}
\usepackage{lineno}
\usepackage{hyperref}
\hypersetup{colorlinks,linkcolor={[RGB]{0 0 139}},citecolor={[RGB]{0 0 139}},urlcolor={[RGB]{0 0 139}}}
\usepackage{amsmath}
\usepackage{subfigure}
\usepackage{multicol}
\usepackage{wrapfig}
\usepackage{titlesec}
\usepackage{array}
\usepackage[dvipsnames]{xcolor}

\usepackage{soul}  
\usepackage{multirow}
\newcommand{\volume}{{\ooalign{\hfil$V$\hfil\cr\kern0.08em--\hfil\cr}}}  

\makeatletter
\def\ps@pprintTitle{%
	\let\@oddhead\@empty
	\let\@evenhead\@empty
	\let\@oddfoot\@empty
	\let\@evenfoot\@oddfoot
}
\makeatother
\journal{*************}

\begin{document}
    \begin{frontmatter}
	\title{Dilatation-driven spurious dissipation in weakly compressible methods}
	\author[1]{Dheeraj Raghunathan}
	\address[1]{School of Mechanical Sciences\\ Indian Institute of Technology Goa\\ Farmagudi-403401, India}
	\author[1]{Y. Sudhakar \corref{cor1}}
	\cortext[cor1]{Corresponding author at: School of Mechanical Sciences, Indian Institute of Technology Goa, Farmagudi - 403401, Goa, India. Email: sudhakar@iitgoa.ac.in}

    \begin{abstract}
The weakly compressible methods to simulate incompressible flows are in a state of rapid development, owing to the envisaged efficiency they offer for parallel computing. The pressure waves in such methods travel at finite speeds, and hence they yield non-solenoidal velocity fields. This inherent inability to satisfy mass conservation corresponding to incompressible flows is a crucial concern for weakly compressible methods. Another widely reported observation is the progressive enhancement of non-physical dissipation with the increase in the artificial compressibility parameter. By scrutinizing the dilatation terms appearing in the kinetic energy equation, we provide vital insights into the influence of mass conservation error on the accuracy of these methods, and explain the mechanism behind the dissipative nature of the compressibility. Analysing transient laminar and turbulent flows, we show that the dilatation-driven dissipation terms, not the mass conservation error alone, govern the accuracy of weakly compressible methods.
The insights provided in this work are not only of fundamental importance but will be of considerable value in aiding the development of weakly compressible methods that can allow a larger artificial Mach number, thus alleviating the stringent time step restriction in such methods.

    \end{abstract}
		

    \end{frontmatter}
	

\section{Introduction}
The \textit{artificial compressibility method} (ACM), proposed by Chorin~\cite{Chorin1967}, does not require solving the numerically expensive pressure Poisson equation associated with widely used computational schemes for incompressible flows. This method, designed for steady-state problems, introduces a pseudo-time derivative of pressure to the continuity equation, thereby adds a slight degree of compressibility to the flow field. 

To enable the simulation of unsteady flows using the ACM, a dual time stepping approach is proposed~\cite{merkle1987}. In this technique, both the continuity and momentum equations are appended with a pseudo-time derivative term. As a consequence, for each physical time step, inner iterations are performed until the pseudo-time derivatives are reduced to a predefined tolerance. Thus, the effect of the artificially added pseudo-time derivatives vanishes, and the procedure is equivalent to solving the unsteady incompressible Navier-Stokes equations. However, the presence of the inner iterations significantly increases the computational cost of dual time stepping ACMs~(DualACM). Nevertheless, this method is utilised extensively for various flow problems~\cite{soh1987,rogers1991upwind,Kiris2002, kim1999, malan2002,loppi2018,ranjan2020}.

In recent years, the development of numerical methods, which introduce compressibility effects artificially to simulate incompressible fluid flows, has taken a new direction. These efforts led to the development of computational approaches that solve, instead of the pressure Poisson equation, an evolution equation for pressure (or a related quantity). Three such computational frameworks that received much attention are the kinetically reduced local Navier Stokes equations~\cite{Karlin2006}, the entropically damped form of artificial compressibility~\cite{Clausen2013} and the general pressure equation~(GPE)~\cite{Toutant2017}. These formulations belong to the class of \textit{weakly compressible methods} (WCM). In addition to these, some of the earlier works in the field of oceanography employed ACM but dropped pseudo-time iterations, as can be inferred from~\cite{madsen2006} and references therein. Recent works~\cite{nagata2021,ikegaya2023} proposed an explicit WCM with virtual particles.

\subsection{Artificial compressibility methods vs Weakly compressible methods}
While ACMs and WCMs mentioned above share the common idea of adding pseudo-compressibility to deal with incompressible flows, they exhibit significant differences as listed below.
\begin{itemize}
    \item The addition of pseudo-time derivatives in ACMs is motivated solely by numerical considerations. However, the pressure evolution equations associated with the WCMs listed above are derived from conservation laws. For example, EDAC~\cite{Clausen2013} is derived from entropy balance equation supported by thermodynamic constitutive relations, and GPE~\cite{Toutant2017} is obtained by modifying the energy conservation equation.
    \item In ACMs, the effect of pseudo compressibility vanishes when the final converged solution is reached. This implies that the artificial Mach number~($Ma$), which quantifies the amount of compressibility introduced in the simulation, affects only the rate of convergence, but has no effect on the converged solution. On the contrary, in WCMs, the compressibility effect does not vanish; hence, $Ma$ plays a crucial role in dictating the accuracy of the solution obtained from WCMs.
    \item The most important difference is that since the pseudo-time derivatives vanish in ACMs, employing such methods is equivalent to solving the incompressible Navier-Stokes equations. The resulting solution accurately satisfies both mass and momentum conservation. This feature is valid for both steady and unsteady flows. However, in WCMs, the continuity equation is replaced with the pressure evolution equation. Since some amount of compressibility is inherent to such an equation, the flowfields obtained from WCMs do not satisfy mass conservation accurately. The severity of the departure from the conservation law depends on the chosen value of $Ma$: the larger the $Ma$, the larger the mass conservation error.
\end{itemize}

The aforementioned arguments highlight the major drawback of WCMs:  they yield velocity fields that exhibit non-zero dilatation i.e., $\nabla\cdot\textbf{u}\neq 0$, which is what we refer to as the mass conservation error in this paper.  
A common notion prevalent among the developers and users of numerical methods for fluid flows is that to obtain physically correct results, the conservation laws must be accurately satisfied. Although WCMs do not strictly obey mass conservation, these methods are becoming increasingly popular because they promise to offer high scalability in parallel computing architectures. Such methods have been used to simulate laminar~\cite{Borok2007,hashimoto2013,Clausen2013,Toutant2018,pan2022}, turbulent~\cite{Shi2020,kajzer2018,vermeire2024,Dupuy2020,trojak2022}, two phase~\cite{kajzer2020,huang2020arxiv,bodhanwalla2024}, buoyancy-driven~\cite{sharma2023}, moving and deforming boundary~\cite{bolduc2023} flows, also on dynamically distorting grids~\cite{abdulgafoor2024}. Moreover, these methods have been applied to develop mesh-free numerical methods to simulate incompressible flows~\cite{ramachandran2019,singh2023}.

\subsection{Contributions of the present work}
Despite the rapidly growing interest in employing WCMs to simulate incompressible flows, the fundamental understanding of such methods is far from complete. The present work aims to provide an improved understanding of WCMs, as underlined in the following two contributions.
\begin{enumerate}
    \item \textit{Influence of mass conservation error:} A crucial concern associated with a WCM is its inherent inability to satisfy mass conservation corresponding to incompressible flows. Although some existing studies on GPE~\cite{Toutant2018,Dupuy2020,pan2022,bodhanwalla2024,raghunathan2024} and EDAC~\cite{Clausen2013,kajzer2018,kajzer2020,vermeire2024,abdulgafoor2024} reported non-negligible mass conservation error produced by these methods, a detailed discussion on the relationship between the error and the accuracy of such approaches has not been deliberated yet in the literature. The present work provides unique insights into the influence of mass conservation error on the accuracy of WCMs. Even in the presence of a significant departure from the divergence-free velocity field, such as $\nabla\cdot\textbf{u}>5$, which is considered to be inadmissible for a simulation, we demonstrate that WCMs yield accurate flowfields representing an incompressible flow. Such a large error and its influence on the accuracy of WCMs have not been reported in the literature, neither for EDAC nor for GPE.
    \item \textit{Dilatation-induced spurious dissipation:} Several works on WCMs~\cite{Clausen2013,Toutant2018,bolduc2023,sharma2023} reported enhanced dissipation when $Ma$ is increased to 0.2 or above. However, the reason for this observation has hitherto not been explained. This, together with the above point, accounts for the following two essential but long-standing questions about WCMs. How do WCMs capture accurate flow fields, even for unsteady configurations, despite their non-negligible mass conservation error? What causes spurious dissipation with the increase in $Ma$? By scrutinizing the kinetic energy evolution equation for WCMs, we identify dilatation terms and answer these important fundamental questions in this paper. This is the key contribution of the present work, and to the best of our knowledge, such objectives have not been attempted in the previous works.
\end{enumerate}

In this paper, we utilized the GPE-based WCM for all our investigations. 
We used DualACM to generate reference incompressible flow solutions. This is motivated by a unique feature of the DualACM that it can provide results corresponding to a pre-specified error in mass conservation. When DualACM is found to be very expensive for a test case, we compare the solutions of the present GPE solver with the reference results published in the literature, or with the results obtained from the popular open-source software OpenFOAM v2012.


\subsection{Structure of the paper}
The rest of the paper is structured as follows. The governing equations and numerical schemes of the GPE and DualACM are presented in section~\ref{section_numMethods}. Then, in \S~\ref{sec:energetics}, we discuss the derivation and interpretation of the kinetic energy equation, which is the crux of the paper. Using numerical examples presented in \S~\ref{test_cases}, we discuss the effects of both mass conservation error and dilatation terms. In section~\ref{salientPoints}, we note down some salient points. Finally, the conclusions are presented in \S~\ref{conclusion_section}.
	
    \section{Governing equations and numerical methods}    
    \label{section_numMethods}
    In this section, the governing equations of both GPE and DualACM solvers are presented, along with the numerical discretisation details. 
	\subsection{General pressure equation (GPE solver) }
	The GPE derived by Toutant~\cite{Toutant2017} is presented below in its non-dimensional form
	\begin{equation}
    	\frac{\partial p}{\partial t} +\frac{1}{Ma^2} \nabla \cdot \mathbf{u} = \frac{\gamma}{RePr} \nabla^{2} p,
    	\label{gpe}
	\end{equation}
        where, $\mathbf{u} = [\,u \quad  v\, \quad w\,]^\top$ is the velocity vector ($u$, $v$ and $w$ being its components in Cartesian coordinate system) and $p$ denotes pressure. $Re$ represents the Reynolds number. The artificial parameters $Ma$, $Pr$ and $\gamma$ indicate the Mach number, Prandtl number and heat capacity ratio, respectively.
	
	The GPE is solved along with the momentum equations (given below) to simulate incompressible flows
	\begin{equation}
    	\frac{\partial \mathbf{u}}{\partial t}  +\nabla\cdot( \mathbf{u}\otimes  \mathbf{u}) = - \nabla p + \frac{1}{Re}\nabla^2 \mathbf{u}.
    	\label{momentum_NS}
	\end{equation}
	Both these equations involve only the physical time derivatives, and hence, no extra set of inner iterations is required.
	\subsection{Dual time stepping artificial compressibility method (DualACM solver)} 
        \label{DualACM_section}
        The artificial compressibility method for simulating unsteady incompressible flows involves solving the following pressure and momentum equations
	\begin{subequations}
	    \begin{equation}
	           \frac{\partial p}{\partial \tau} + \frac{1}{Ma^2}  \nabla \cdot \mathbf{u} = 0
	           \label{pressure_chorin}
	    \end{equation}
            \begin{equation}
                \frac{\partial \mathbf{u}}{\partial \tau} + \frac{\partial \mathbf{u}}{\partial t} + \nabla\cdot( \mathbf{u}\otimes  \mathbf{u}) = - \nabla p + \frac{1}{Re} \nabla^2 \mathbf{u},
    	    \label{dual_time_stepping_momentum}
            \end{equation}
	\end{subequations}
        where, $\mathbf{u}$, $p$, $Re$ and $Ma$ denotes the same parameters described above. This formulation requires marching in both physical time~($t$) and pseudo-time~($\tau$), and hence, referred to as the dual time stepping scheme. At each physical time advancement, we perform time marching in $\tau$ until the pseudo-time derivatives reduce to a preset tolerance value. i.e., $\partial \mathbf{u}/\partial \tau < [\epsilon_{mom}^x \quad  \epsilon_{mom}^y \quad \epsilon_{mom}^z]^\top$ and $Ma^2\partial p/\partial \tau < \epsilon_{p}$. Hence, the DualACM is equivalent to solving the incompressible Navier-Stokes equations at all time instants; the effect of artificial compressibility, introduced by $Ma$, does not affect the solution at any time step. 
        In all the simulations that we discuss later in section~\ref{test_cases}, we set, $\epsilon_{mom}^x = \epsilon_{mom}^y = \epsilon_{mom}^z = \epsilon_{p} = \epsilon$.  
	\subsection{Discretization}
	\begin{figure}[!ht]
		\centering
		\includegraphics[trim = 0mm 0mm 0mm 0mm, clip, width=15cm]{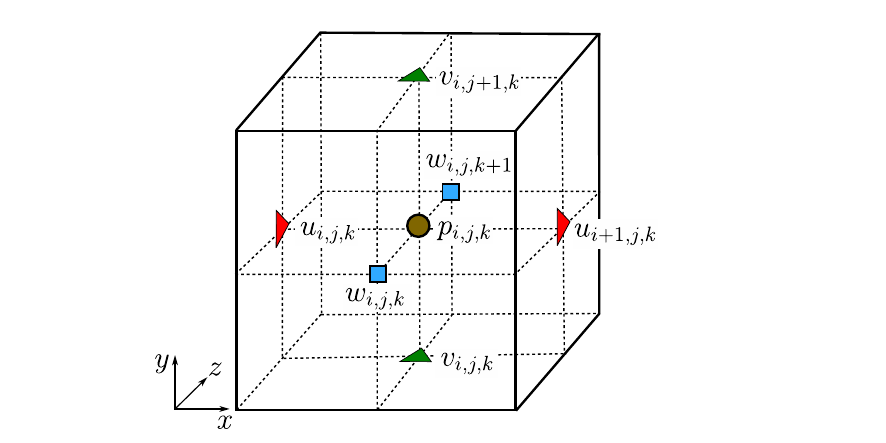}
		\caption{The staggered grid arrangement used for the discretization of the governing equations.}
		\label{staggered}
	\end{figure}
        The governing equations are written in their integral form, and a finite volume approach is adopted to discretize the governing equations of both GPE and DualACM. We use the Cartesian coordinate system, and the computational stencil is illustrated in figure~\ref{staggered}. To avoid pressure-velocity decoupling, we use a staggered grid arrangement, where the pressure, $p$, is stored at the cell centre, and the velocity components $u$, $v$, and $w$ at the cell face centres. The convective terms, pressure gradient terms and diffusion terms in the equations are approximated using the central difference scheme similar to~\cite{Toutant2018,Dupuy2020}. A representative example of discretisation in a uniform Cartesian grid is provided below. All the other terms follow a similar approach.
 
 	\begin{align*}
    	\int_\volume \nabla \cdot (u \mathbf{u}) d\volume \approx
    	& \biggl[ \left( \frac{u_{i+1,j,k} + u_{i,j,k}}{2} \right)^2  - 
    	\left( \frac{u_{i,j,k} + u_{i-1,j,k}}{2} \right)^2 \biggr] \Delta y \Delta z \quad + \\
    	& \biggl[ \left( \frac{u_{i,j+1,k} + u_{i,j,k}}{2} \right)
    	\left( \frac{v_{i-1,j+1,k} + v_{i,j+1,k}}{2} \right) \quad - \\
    	& \ \left( \frac{u_{i,j,k} + u_{i,j-1,k}}{2} \right) 
    	\left( \frac{v_{i-1,j,k} + v_{i,j,k}}{2} \right) \biggr] \Delta x \Delta z \quad + \\
    	& \biggl[ \left( \frac{u_{i,j,k+1} + u_{i,j,k}}{2} \right)
    	\left( \frac{w_{i-1,j,k+1} + w_{i,j,k+1}}{2} \right) \quad - \\
    	& \ \left( \frac{u_{i,j,k} + u_{i,j,k-1}}{2} \right) 
    	\left( \frac{w_{i-1,j,k} + w_{i,j,k}}{2} \right) \biggr] \Delta x  \Delta y
	\end{align*}

	\begin{align*}
    	\int_\volume \frac{\partial p}{\partial x} d\volume 
    	&\approx  \left( \frac{p_{i,j,k} - p_{i-1,j,k}}{\Delta x} \right) \Delta x \Delta y \Delta z
	\end{align*}
 	
	\begin{align*}
    	\int_\volume \ \nabla \cdot (\nabla u) d\volume \approx
    	& \biggl[ \left( \frac{u_{i+1,j,k} - u_{i,j,k}}{\Delta x } \right) - 
    	\left( \frac{u_{i,j,k} - u_{i-1,j,k}}{\Delta x} \right) \biggr] \Delta y \Delta z \quad + \\
    	& \biggl[ \left( \frac{u_{i,j+1,k} - u_{i,j,k}}{\Delta y} \right)  - 
    	\left( \frac{u_{i,j,k} - u_{i,j-1,k}}{\Delta y } \right) \biggr]  \Delta x \Delta z \quad + \\
    	& \biggl[ \left( \frac{u_{i,j,k+1} - u_{i,j,k}}{\Delta z} \right)  - 
    	\left( \frac{u_{i,j,k} - u_{i,j,k-1}}{\Delta z} \right) \biggr] \Delta x  \Delta y
	\end{align*}
        Here, $i$, $j$ and $k$ indicate the indices and $\Delta x$, $\Delta y$ and $\Delta z$ represent the grid size in $x$, $y$ and $z$ directions respectively. Note that, the convective and diffusion terms above utilised the Gauss divergence theorem. 

        For temporal discretisation of DualACM equations, we adopt the approach used by Parameswaran and Mandal~\cite{Parameswaran2019}, where the two time derivatives (based on $t$ and $\tau$) are discretized separately. A three-stage Strong-Stability Preserving Runge-Kutta (SSP-RK) method is used to discretize the pseudo-time derivative, and a three-point implicit backward difference formula is used for the physical time derivative. This discretisation procedure is briefly explained below. 

	From equation \eqref{dual_time_stepping_momentum}, the pseudo-time derivative can be rewritten as follows.
	\begin{equation}
    	\frac{\partial \mathbf{u}}{\partial \tau}  = -\frac{\partial \mathbf{u}}{\partial t}  + L(\mathbf{u})
    	\label{dualacm_momentum_modified}
	\end{equation}
	where $L$ is the spatial discretisation operator. The physical time derivative term is approximated by,
	\begin{align}
    	\frac{\partial \mathbf{u}}{\partial t} &= \begin{cases}
    	\dfrac{\mathbf{u}^{(k)} -\mathbf{u}^{(0)} }{ \Delta t} & \text{if, $t=0$}  \\  \\
    	\dfrac{\mathbf{u}^{(k)} -4\mathbf{u}^{(n)} + \mathbf{u}^{(n-1)}  }{2 \Delta t} & \text{if, $t>0$}
    	\end{cases}
    	\label{implicit_BD}
	\end{align}
        where superscripts $n$ and $n-1$ represent the current and previous time levels. $\mathbf{u}^{(0)}$ represents the initial condition. Also, $k$ denotes an intermediate pseudo-time level, and $\Delta t$ is the physical time step. 
	To update $\mathbf{u}$ from $k$ to $k+1$, we follow the steps given below.
	
	\begin{linenomath}
		\begin{align}
    		\begin{split}
        		\mathbf{u}^{(1)} =& \:\:\: \mathbf{u}^{(k)} + \frac{\Delta \tau}{\volume} \textbf{R}(\mathbf{u}^{(k)}) \\
        		\mathbf{u}^{(2)} =& \:\:\: \frac{3}{4}\mathbf{u}^{(k)} + \frac{1}{4}\mathbf{u}^{(1)} + \frac{1}{4} \frac{\Delta \tau}{\volume} \textbf{R}(\mathbf{u}^{(1)}) \\
        		\mathbf{u}^{(k+1)} =& \:\:\: \frac{1}{3}\mathbf{u}^{(k)} + \frac{2}{3}\mathbf{u}^{(2)} + \frac{2}{3} \frac{\Delta \tau}{\volume} \textbf{R}(\mathbf{u}^{(2)})  
    		\end{split}
                \label{sspRK}
		\end{align}
	\end{linenomath}
        Here, \textbf{R} denotes the RHS of equation \eqref{dualacm_momentum_modified}, which includes both the discretized spatial derivatives and physical time derivative (approximated using equation \eqref{implicit_BD}). $\Delta \tau$ is the pseudo-time step.
	
        For a fair comparison between DualACM and GPE, we use the temporal discretisation method specified by equation~\eqref{sspRK} for the time integration of the GPE solver. 
 

         For GPE, the requirement of $Ma \ll 1$ introduces a stringent stability limit on the time step. In most cases, the time step is determined by the Courant–Friedrichs–Lewy~(CFL) condition defined based on the artificial acoustic wave speed. This also applies to the DualACM but for the pseudo-time step. Hence, for the same $\textrm{Ma}$, we have $\Delta \tau_{DualACM} = \Delta t_{GPE}$. A detailed discussion on the choice of time step for the GPE solver is presented by Raghunathan and Sudhakar~\cite{raghunathan2024}. It is well-known~\cite{gaitonde1998,Parameswaran2019} that for stable computation using the DualACM scheme, the pseudo-time derivative $\Delta \tau \leq 2\Delta t/3$. Complying with this, we fix $\Delta \tau_{DualACM} = \Delta t_{DualACM}/2$ for all the simulations that are presented in the following section.

\section{Energetics of the flow}
\label{sec:energetics}

The building block of the discussions presented in this paper is the equation for kinetic energy. In this section, we briefly derive and discuss the physical interpretation of all terms in the equation.

We start with the momentum equation, written in indicial notation
\begin{equation}
    \frac{\partial u_i}{\partial t}+\frac{\partial u_iu_j}{\partial x_j}=-\frac{\partial p}{\partial x_i}+\frac{1}{Re}\frac{\partial^2 u_i}{\partial x_j\partial x_j}.
\end{equation}
In order to derive the evolution equation for kinetic energy, $k=u_iu_i/2$, we multiply the above equation by $u_i$. After the multiplication and straightforward manipulations we get
\begin{equation}
\frac{\partial \left(\frac{1}{2}u_iu_i\right)}{\partial t}+u_j\frac{\partial }{\partial x_j}\left(\frac{1}{2}u_iu_i\right)+u_iu_i\frac{\partial u_j}{\partial x_j}=-u_i\frac{\partial p}{\partial x_i}+\frac{1}{Re}u_i\frac{\partial^2 u_i}{\partial x_j\partial x_j}.
\end{equation}
We employ the following two modifications to the above equations: (i)~introduce the well-known material derivative~($D/Dt$), and (ii)~apply the Leibniz product rule of the calculus and collect all the terms that involve the divergence operator. After these modifications, the equation becomes,
\begin{equation}
    \frac{Dk}{Dt}=\frac{\partial}{\partial x_j}\left(-pu_j+\frac{1}{Re}u_i\frac{\partial u_i}{\partial x_j}\right)-u_iu_i\frac{\partial u_j}{\partial x_j}+p\frac{\partial u_i}{\partial x_i}-\frac{1}{Re}\frac{\partial u_i}{\partial x_j}\frac{\partial u_i}{\partial x_j}.
\end{equation}

We integrate the above differential equation over the domain $\Omega$ whose volume is $ \volume$, and the bounding surfaces of the domain are denoted by $\Gamma$. Application of the divergence theorem for the first term on the right-hand side yields,
\begin{align*}
    \frac{D}{Dt}\int_ \volume {k d\volume}
    &=\int_\Gamma{\left(-pu_j+\frac{1}{Re}u_i\frac{\partial u_i}{\partial x_j}\right)n_j}d\Gamma 
    -\int_ \volume{u_iu_i\frac{\partial u_j}{\partial x_j}d \volume} \\
    &-\int_ \volume{\left(-p\frac{\partial u_i}{\partial x_i}\right)d \volume}
    -\int_ \volume{\frac{1}{Re}\frac{\partial u_i}{\partial x_j}\frac{\partial u_i}{\partial x_j}d \volume}.
\end{align*}
For all the numerical examples considered in this paper, we enforce either the periodic condition or zero Dirichlet condition for velocity components on $\Gamma$. Hence, the surface integral term in the above equation, which represents the external work done at the boundaries, goes to zero. Thus, the kinetic energy equation in integral form is simplified to
\begin{equation}
    \frac{D}{Dt}\int_ \volume {k d\volume}=-\mathcal{D}_k\volume-\mathcal{D}_p\volume-E_\Delta\volume,
    \label{eqn:ke_eqn}
\end{equation}
where the left-hand side quantifies the rate of change of total kinetic energy in the domain, $\mathcal{D}_k$ and $\mathcal{D}_p$ denote (average) kinetic energy dilatation and (average) pressure dilatation, respectively. They are defined as
\begin{align}
    \mathcal{D}_k&=\frac{1}{\volume}\int_ \volume{u_iu_i\frac{\partial u_j}{\partial x_j}d \volume}\label{eqn:ke_dila}\\
    \mathcal{D}_p&=\frac{1}{\volume}\int_ \volume{\left(-p\frac{\partial u_i}{\partial x_i}\right)d \volume}\label{eqn:p_dila}.
\end{align}
Both these terms stem from the non-solenoidal nature of the velocity field in the weakly compressible methods. For an incompressible flow, both terms are zero ($\mathcal{D}_k=\mathcal{D}_p=0$) since the velocity field is solenoidal. Hence, for an incompressible flow, in the absence of external interaction, the kinetic energy decays due to the viscous forces~\cite{mathieu2000}. However, the presence of $\mathcal{D}_k$ and $\mathcal{D}_p$ in weakly compressible methods adds additional features. In the upcoming sections, we will elaborate that these crucial dilatation terms dictate the accuracy of weakly compressible methods, not the mass conservation error alone. Moreover, we will show that these terms add spurious dissipation, and hence are responsible for enhanced dissipation when $Ma$ is increased.

Research works on compressible turbulence \cite{sarkar1991,sarkar1992,miura1995,mittal2020,wang2021} extensively investigated some form of the pressure dilatation term on the turbulence dynamics. However, they were concerned with the mean dilatation term (defined as the product of pressure fluctuations and dilatational fluctuations), and its effect on the turbulent kinetic energy and empirical modelling. The arguments presented in these references are not directly relevant to the present work. So they are not reviewed in detail here. 

The last term of equation~\eqref{eqn:ke_eqn} is the  average viscous dissipation rate of kinetic energy
\begin{equation}
    E_\Delta = \frac{1}{\volume}\int_ \volume{\frac{1}{Re}\frac{\partial u_i}{\partial x_j}\frac{\partial u_i}{\partial x_j}d \volume},
\end{equation}
which is different from the classical version presented in terms of the strain rate. This is because although the velocity field is not divergence-free, it is common in WCMs, to express the viscous terms in momentum equations in terms of the Laplacian of velocity.

For quantitative verification of the numerical examples in the next section, we compute the evolution of the integral quantities such as average kinetic energy, $E_k$ and enstrophy, $\zeta$. For completeness, we present their definition here
	\begin{equation}
    	E_k = \frac{1}{ \volume} \int_\volume  \frac{u_i u_i}{2} d\volume, 
    	\label{ke_equation}
	\end{equation}
	and
	\begin{equation}
    	\zeta = \frac{1}{ \volume} \int_\volume \omega_i\omega_i d\volume, 
    	\label{enstrophy_eqn}
	\end{equation}
where $\omega_i$ denotes the vorticity vector.
    \section{Numerical experiments}
    \label{test_cases}
    We present numerical examples to investigate the effect of the dilatation-driven spurious dissipation terms and mass conservation errors on the accuracy of the present GPE-based weakly compressible approach. We start with a detailed analysis of the two-dimensional doubly periodic shear layer problem to expose the importance of the spurious dissipation terms. This is followed by a rigorous examination of flow structures on the test case involving dipole-wall interaction that incurs exceedingly large mass conservation errors. As a final numerical example, we report the three-dimensional Taylor-Green vortex problem to extend our analysis to the turbulent flow regime.
    
    Similar to the previous works on GPE~\cite{Toutant2018,raghunathan2024}, for all the test cases, we set $\textrm{Pr}=\gamma$. The numerical examples using GPE are simulated with artificial Mach numbers of 0.02, 0.1 and 0.2.
    
    The convergence criterion of $\epsilon=10^{-6}$ is used for the DualACM simulations. This implies the mass conservation error is less than $10^{-6}$ at every point in the domain. Hence, these results can be considered to represent reference incompressible flow simulations, for which the dilatation terms are absent. For the numerical examples that follow, the time step used is listed in table~\ref{timestepTable}. 
    \begin{table}[H]
        \centering
        \caption{Time steps used in the numerical examples.}
        \footnotesize
        \begin{tabular}{lllll}
            \: \\
            \hline
            \hline\\
            \textbf{Test Case} &&  \textbf{GPE} ($\boldsymbol{\Delta t}$) &&  \textbf{DualACM} ($\boldsymbol{\Delta t}$ , $\boldsymbol{\Delta \tau} $) \\
            \hline\\
            Doubly periodic shear layer && $2\times 10^{-5}$ && $4\times 10^{-5}$, $2\times 10^{-5}$  \\
            Dipole-wall collision && $1\times 10^{-5}$ &&  \multicolumn{1}{c}{--}  \\
            turbulent Taylor-Green vortex ($256^3$ mesh) && $2.5\times 10^{-4}$ && $5\times 10^{-4}$, $2.5\times 10^{-4}$  \\
            turbulent Taylor-Green vortex ($512^3$ mesh) && $1\times 10^{-4}$ && \multicolumn{1}{c}{--}  \\
            \hline
            \hline
        \end{tabular}
        \label{timestepTable}
    \end{table}
    
\subsection{Doubly periodic shear layer}
\label{sec:dpsl}
The doubly periodic shear layer (DPSL) test case has been widely used to validate various incompressible flow solvers, including GPE~\cite{Clausen2013,hashimoto2013,hashimoto2015,Toutant2018}. Here, our discussion is focused on how the dilatation terms~($\mathcal{D}_k$ and $\mathcal{D}_p$) appearing in the equation of kinetic energy affect the accuracy of the GPE-based incompressible flow solver.

In this problem, the initial velocity field is characterised by two horizontal shear layers perturbed in the vertical direction. Following Minion and Brown~\cite{minion1997}, we set the initial conditions as follows,
\begin{subequations}
\begin{align}
    u &=  \begin{cases}
    \tanh(\rho (y-0.25)),      & \text{if} \hspace{0.2cm} {y \le 0.5} \\
    \tanh(\rho (0.75-y)),      & \text{otherwise}
    \end{cases} \\
    v &= \delta \sin (2 \pi (x + 0.25)) \\
    p &= 0
\end{align}
\end{subequations}
where $\rho$ is the shear layer thickness and $\delta$ represents the amplitude of the initial perturbation. We use $\rho=80$ and $\delta = 0.05$, which are  widely used by other  researchers~\cite{Clausen2013,hashimoto2013,hashimoto2015,Toutant2018}. The computational domain is a periodic unit square [$1 \times 1$], divided into a uniform $512 \times 512$ grid. The Reynolds number, $Re=10^{4}$. The simulation is run until time $t=5$.

\begin{figure}[!ht]
\begin{center}
    \includegraphics[trim = 0mm 0mm 0mm 0mm, clip, width=15cm]{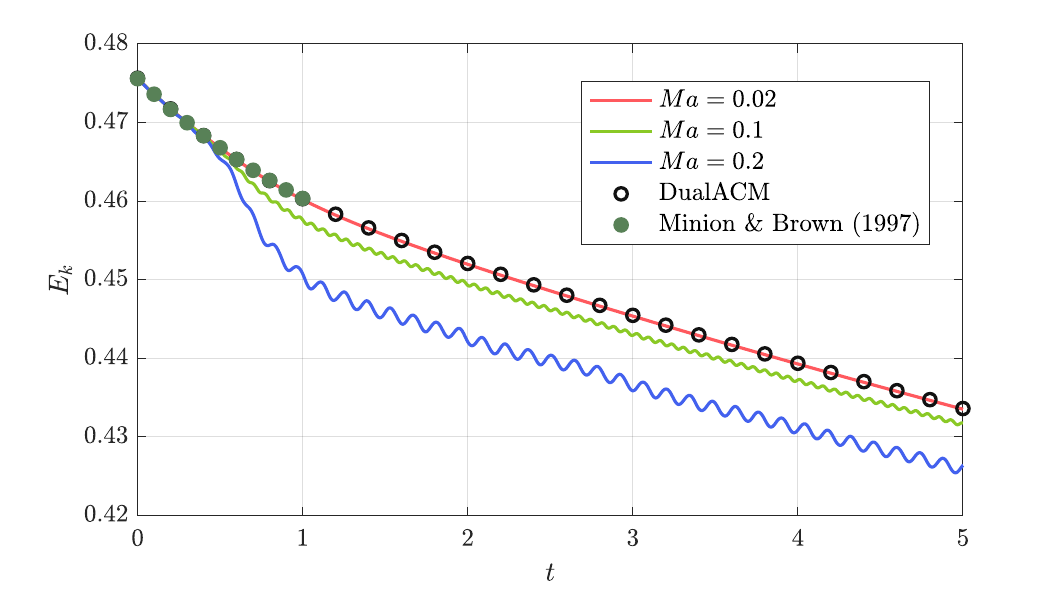}
\end{center}
\caption{Evolution of average kinetic energy obtained from GPE and DualACM schemes for the doubly periodic shear layer test case. Reference values reported by Minion and Brown~\cite{minion1997} are also included.}
\label{fig:Dpsl_ke}
\end{figure} 

The temporal evolution of the average kinetic energy for different $Ma$ is plotted in figure~\ref{fig:Dpsl_ke}. Results from DualACM with $\epsilon=10^{-6}$ are also included. Moreover, these transient plots are compared with the reference data extracted from Minion and Brown~\cite{minion1997}. It is directly evident from the figure that when $Ma=0.02$, the kinetic energy is in excellent agreement with DualACM and the reference results. However, as $Ma$ is increased, the accuracy of the simulation progressively deteriorates. 
       
\begin{figure}[!ht]
\centering
\subfigure[]
{
\includegraphics[trim = 20mm 0mm 20mm 0mm, clip, width=5.1cm]{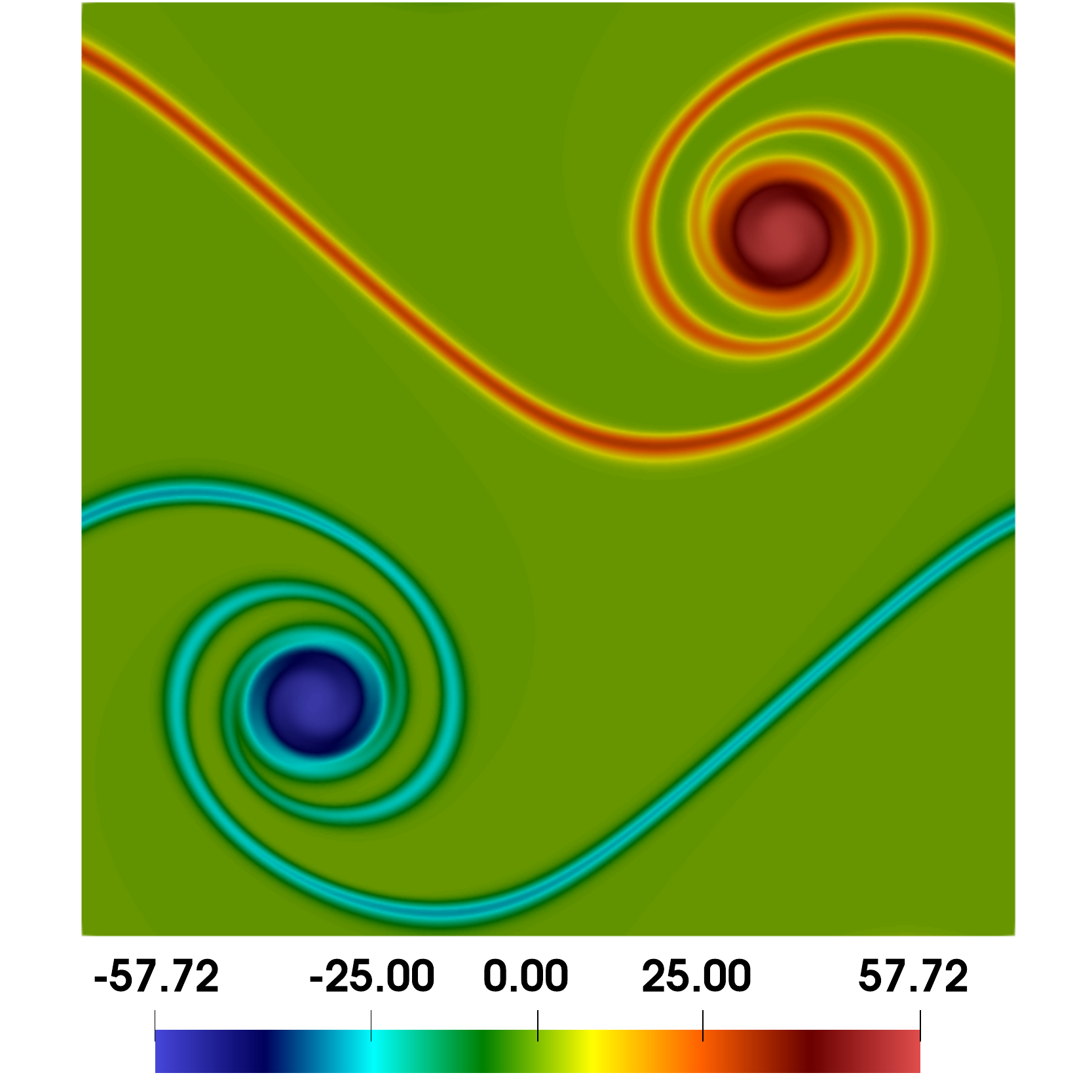}
\label{dpsl_vort_comp2}
}
\subfigure[]
{
\includegraphics[trim = 20mm 0mm 20mm 0mm, clip, width=5.1cm]{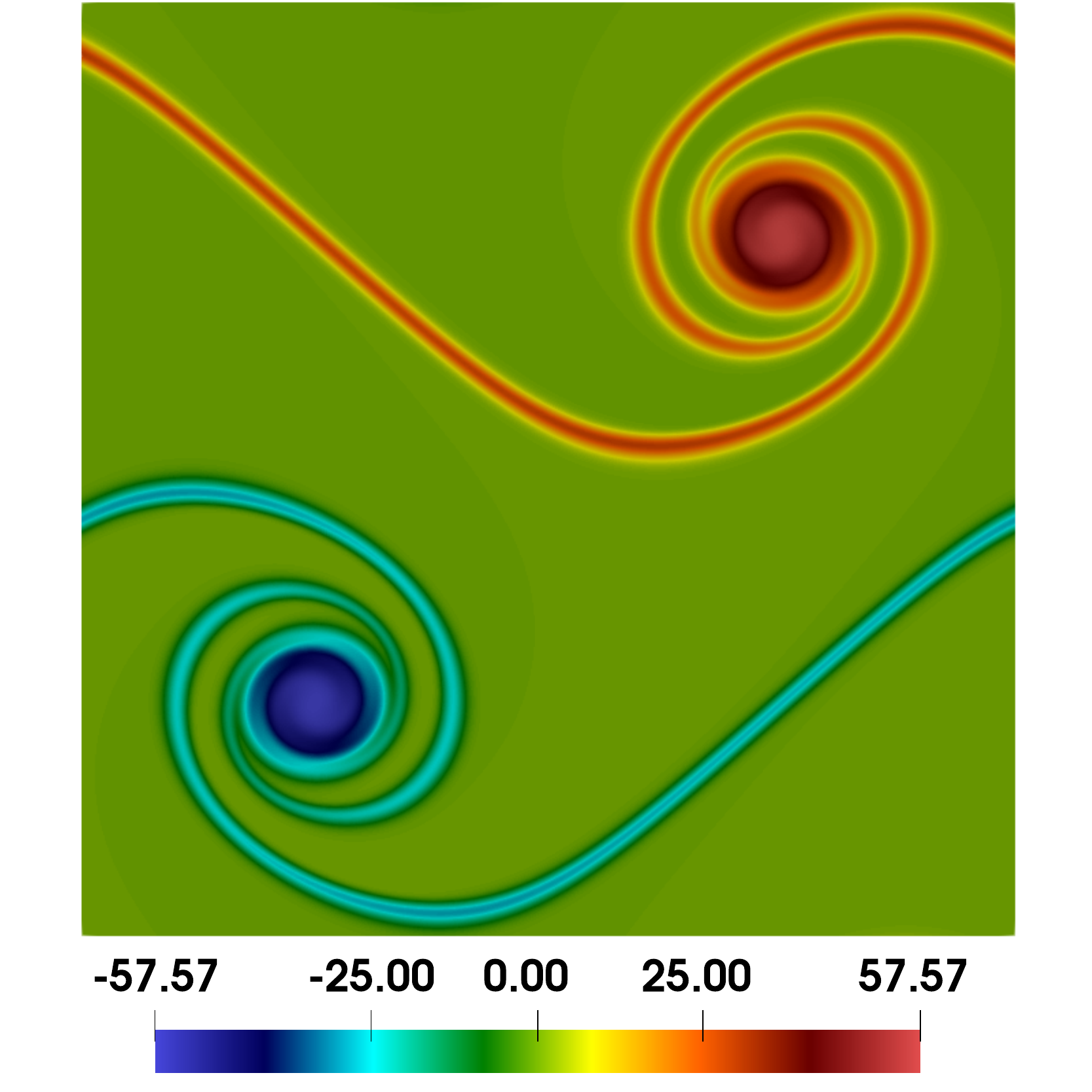}
\label{dpsl_vort_comp5}
}
\subfigure[]
{
\includegraphics[trim = 20mm 0mm 20mm 0mm, clip, width=5.1cm]{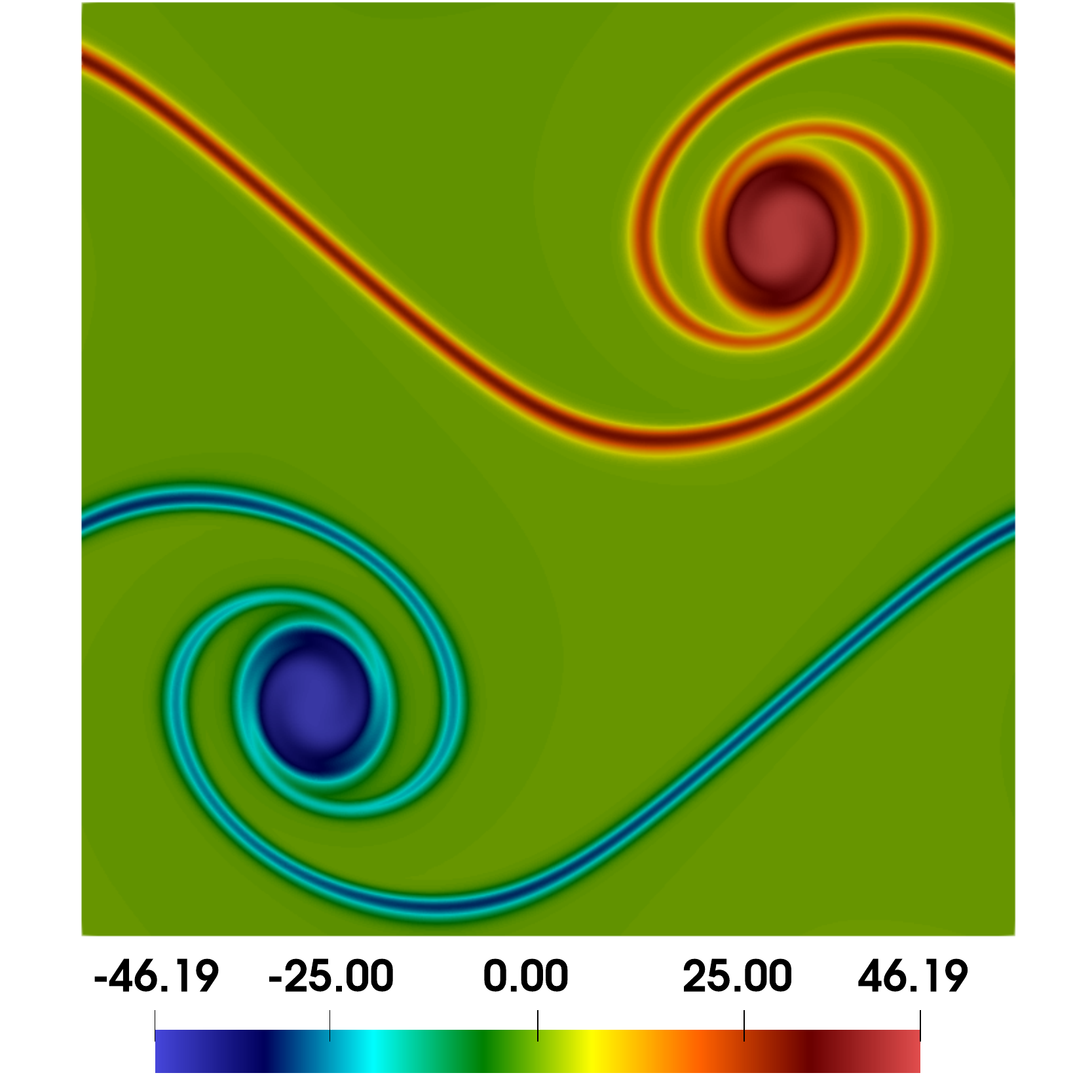}
\label{dpsl_vort_comp512}
}
\caption{Contours of $z-$component of vorticity obtained from (a)~DualACM, (b)~GPE with $Ma=0.02$, and (c)~GPE with $Ma=0.2$.}
\label{dpsl_vort_comp}
\end{figure}

Inspection of the vorticity contours at $t=1$, illustrated in figure~\ref{dpsl_vort_comp}, provides additional confirmation for this observation. Results from $Ma=0.02$ and DualACM are identical (as evidenced also in \ref{app:dpsl}). However, at $Ma=0.2$, the magnitude of vorticity is reduced implying enhanced dissipation, consistent with the kinetic energy evolution given in figure~\ref{fig:Dpsl_ke}.

The aforementioned findings are not surprising because as $Ma$ increases, the governing equations of any weakly compressible approach depart from describing the behaviour of incompressible flows. Although such observation for this same test case (until $t=1$) is presented using GPE~\cite{Toutant2018} as well as EDAC~\cite{Clausen2013,bolduc2023,sharma2023}, the \textit{precise} answer to why this happens, especially the reason for increased dissipation observed with $Ma$, is not answered in the literature. In the following, we prove that the dilatation terms appearing in the equation of kinetic energy are the source of this additional spurious dissipation.

\begin{figure}[!ht]
\centering
\subfigure[]
{
    \includegraphics[trim = 0mm 0mm 0mm 0mm, clip, width=7.5cm]{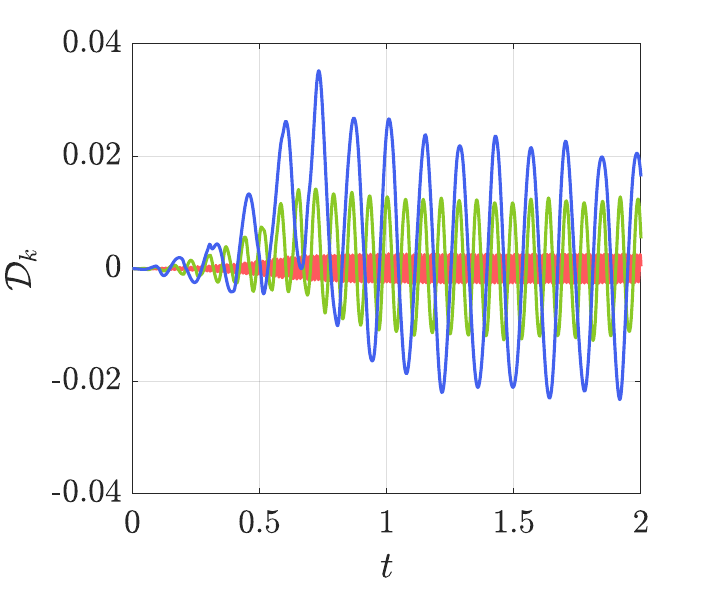}
}
\subfigure[]
{
    \includegraphics[trim = 0mm 0mm 0mm 0mm, clip, width=7.5cm]{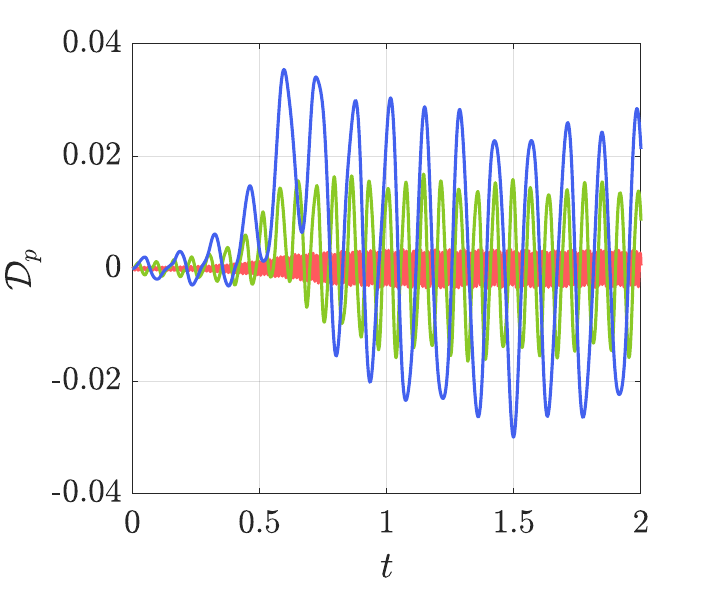}
}
\newline
\subfigure[]
{
    \includegraphics[trim = 0mm 0mm 0mm 0mm, clip, width=7.5cm]{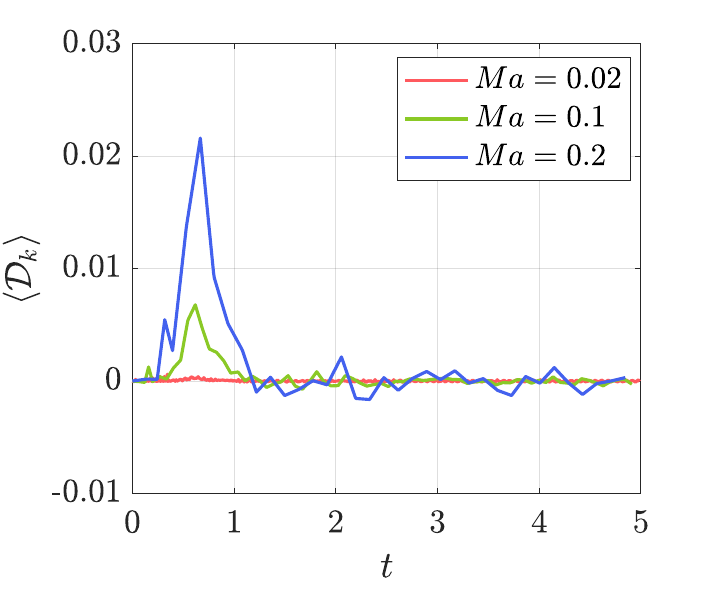}
}
\subfigure[]
{
    \includegraphics[trim = 0mm 0mm 0mm 0mm, clip, width=7.5cm]{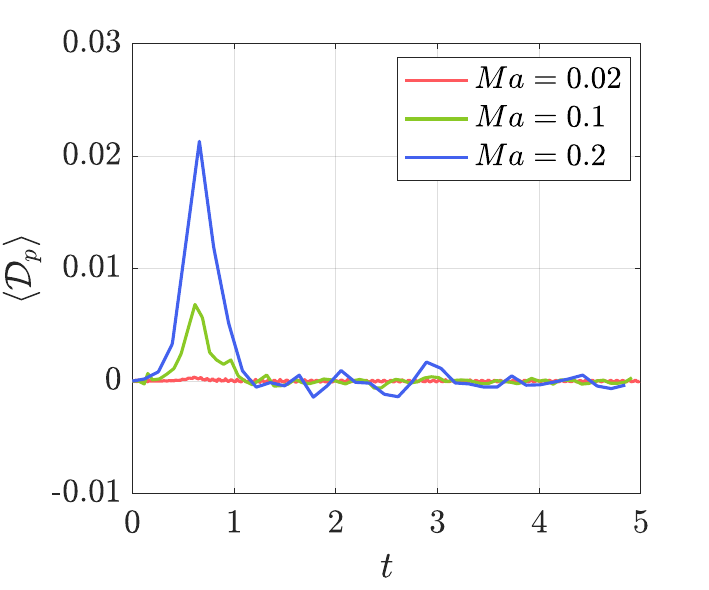}
}
    \caption{Evolution of kinetic energy dilatation and pressure dilatation for the doubly periodic shear layer test case: (a)~$\mathcal{D}_k$ computed using equation~\eqref{eqn:ke_dila}, (b)~$\mathcal{D}_p$ computed using equation~\eqref{eqn:p_dila}, (c)~cycle-averaged $\mathcal{D}_k$,  (d)~cycle-averaged $\mathcal{D}_p$.}
    \label{dpsl_ke_p_dila}
\end{figure}
      
The variation of kinetic energy dilatation and pressure dilatation until $t=2$ is presented in figure~\ref{dpsl_ke_p_dila}(a) and (b), respectively. We can observe that both $\mathcal{D}_k$ and $\mathcal{D}_p$ exhibit oscillations, whose magnitude increases and the frequency reduces with increasing $Ma$. These observations are consistent with the fact that they arise due to the presence of weak compressibility; the strength of artificial acoustic waves increases and the frequency reduces with increasing $Ma$. The oscillations are physically characterized by the energy exchange between kinetic energy and potential mode, analogous to compressible turbulence~\cite{sarkar1991,miura1995,wang2021}.


\begin{figure}[!ht]
\begin{center}
\includegraphics[trim = 0mm 0mm 0mm 0mm, clip, width=15cm]{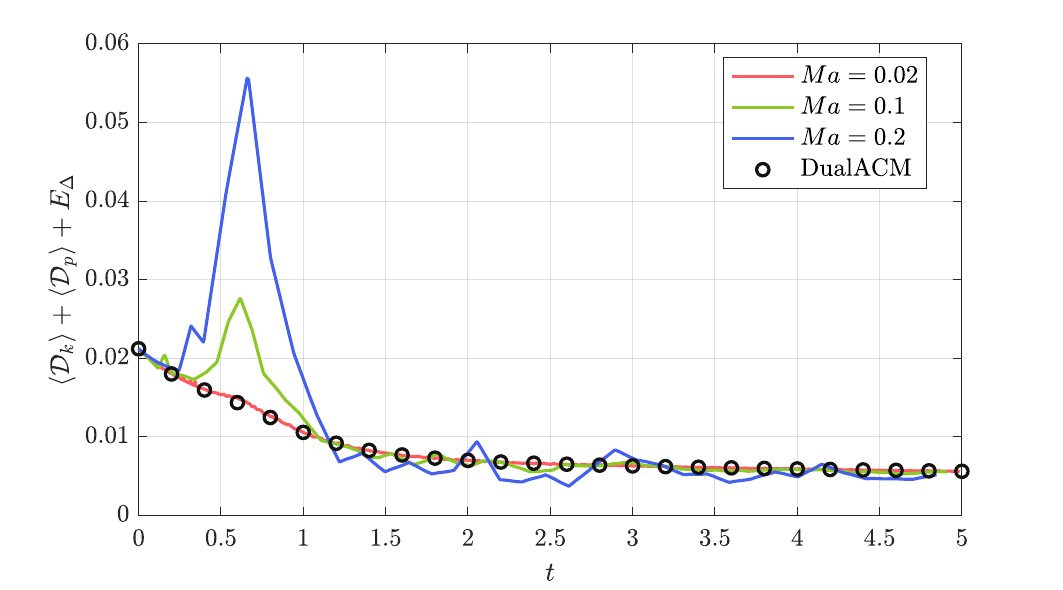}
\end{center} 
\caption{Total kinetic energy dissipate rate per unit volume for the doubly periodic shear layer test case.}
\label{dpsl_dissi}
\end{figure} 

For a better physical interpretation, we compute the cycle-averaged values of these dilatation terms, denoted $\left<\mathcal{D}_k\right>$ and $\left<\mathcal{D}_p\right>$, and present their time evolution in figure~\ref{dpsl_ke_p_dila}(c) and (d), respectively. These values are significant until $t=1$. 
The peak in $\left<\mathcal{D}_k\right>$ and $\left<\mathcal{D}_p\right>$ increases with $Ma$, but the time at which it occurs is not strongly influenced by $Ma$. 
It can be interpreted from equation~\eqref{eqn:ke_eqn} that this leads to an enhanced dissipation of the kinetic energy that is directly evident from figure~\ref{fig:Dpsl_ke}.

\begin{table}[!ht]
\centering
\caption{The contribution of the sum of dilatation terms~($\left<\mathcal{D}_k\right>+\left<\mathcal{D}_p\right>$) to the rate of dissipation. Here $\hat{t}$ denotes the time at which the above sum is maximum for the respective $Ma$, and $E_\Delta$ from DualACM is reported at the corresponding time.}
\footnotesize
\begin{tabular}{cccc}
    \hline
    \hline\\
$Ma$ & $\hat{t}$  & $E_\Delta$ at $t=\hat{t}$ & $\left<\mathcal{D}_k\right>+\left<\mathcal{D}_p\right>$\\ \hline 
0.02 & 0.3429 &  0.0165  &   0.0006 \\
0.1  & 0.6186 &  0.0142  &  0.0136  \\
0.2  & 0.6590 &  0.0138  &  0.0424  \\
    \hline
    \hline
\end{tabular}
\label{tab:dpsl_dissi}
\end{table}

Further insights can be gained by investigating the total kinetic energy dissipation rate per unit volume, $\left<\mathcal{D}_k\right>+\left<\mathcal{D}_p\right>+E_\Delta$, which is the sum of all the terms on the right-hand of the equation~\eqref{eqn:ke_eqn}. The evolution of this quantity, presented in figure~\ref{dpsl_dissi}, reveals that the data from $Ma=0.02$ matches very well with that of the physical dissipation in DualACM for which the dilatation terms are zero. This implies that at such low $Ma$, the dilatation terms' influence can be neglected. However, at $Ma=0.1$ and 0.2, the dilatation terms noticeably pump up the rate of dissipation. We label it as spurious dissipation because this is absent in incompressible flows. This effect is quantified in table~\ref{tab:dpsl_dissi} that presents the time~($\hat{t}$) at which the value of $\left<\mathcal{D}_k\right>+\left<\mathcal{D}_p\right>$ is maximum, the respective value of this spurious dissipation, and the viscous dissipation rate of DualACM at $t=\hat{t}$. At $Ma=0.1$, the dilatation term makes almost equal contribution as $E_\Delta$. However, at $Ma=0.2$, the spurious dissipation rate is more than thrice that of $E_\Delta$, confirming a stronger dominance.

\begin{figure}[!ht]
    \centering
        \includegraphics[trim = 0mm 0mm 0mm 0mm, clip, width=15cm]{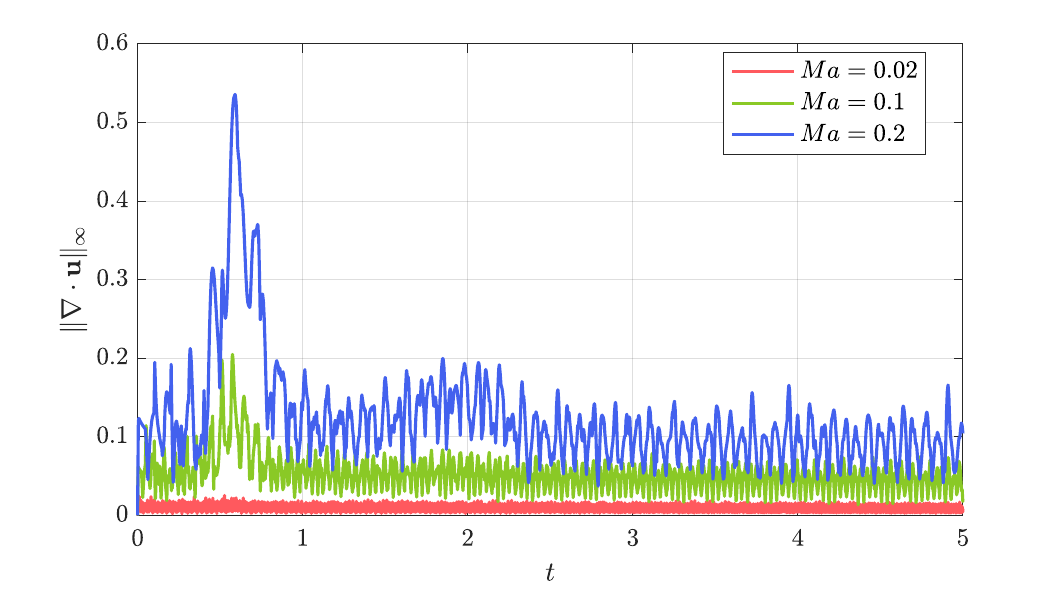}
        \label{dpsl_velDivMax_machStudy}
    \caption{Evolution of $L_\infty-$norm of the mass conservation error for the doubly periodic shear layer problem using $512^2$ mesh.}
    \label{dpsl_velDiv_machStudy}
\end{figure} 

It is directly evident that the peak in the rate of dissipation, as shown in figure~\ref{dpsl_dissi}, corresponds to the rapid departure of kinetic energy from the DualACM, as illustrated in figure~\ref{fig:Dpsl_ke}. This observation proves that dilatation-induced spurious dissipation is the reason behind the enhanced dissipation reported for WCMs at larger $Ma$. Even though previous works reported incorrect estimation of kinetic energy at higher $Ma$, to the best of our knowledge, this is the first time an explanation is provided for the same.

The presence of non-physical oscillations in time for the results obtained from WCMs with large $Ma$ has been reported earlier. For this same problem, many studies~\cite{Clausen2013,Toutant2018,bolduc2023,sharma2023} have shown the oscillations in $E_k$ as shown in figure~\ref{fig:Dpsl_ke}. Moreover, Bodhanwalla et al.~\cite{bodhanwalla2024} have observed oscillations in the shape of the interface in their two-phase flow simulations using GPE. However, the source of these oscillations has not been explored in the literature. The energy exchange mechanism driven by the dilatation terms reported here accounts for these oscillations also.

For completeness, we present the variation of velocity divergence, in terms of $L_\infty-$norm, with time in figure~\ref{dpsl_velDiv_machStudy}. The maximum $\nabla\cdot\textbf{u}$ is of the order of $0.5$ for $Ma=0.2$. Such a large error naturally casts strong suspicion that this might be the root cause of the inaccuracy. However, the next test case shows that even larger values of $\nabla\cdot\textbf{u}$ can leave the flowfield unaffected if the dilatation-driven dissipation terms are insignificant.
\subsection{Dipole-wall collision}
\label{sec:dipole}
In order to study the role of the spurious dissipation induced by the dilatation in wall-bounded flows, we simulate a challenging test case in which a vortex dipole collides with a wall. This numerical example involves the creation and separation of thin boundary layers that contain high-amplitude vorticity patches during the collision process. The ensuing flow is highly dependent on the small-scale vortical structures produced near the wall, whose dynamics must be correctly captured for accurate computation of the flow field. Therefore, this problem serves as a particularly rigorous test case for numerical algorithms~\cite{clercx2006,keetels2007}.
        
\begin{figure}[!ht]
\centering
\includegraphics[trim = 0mm 0mm 0mm 0mm, clip, width=7.5cm]{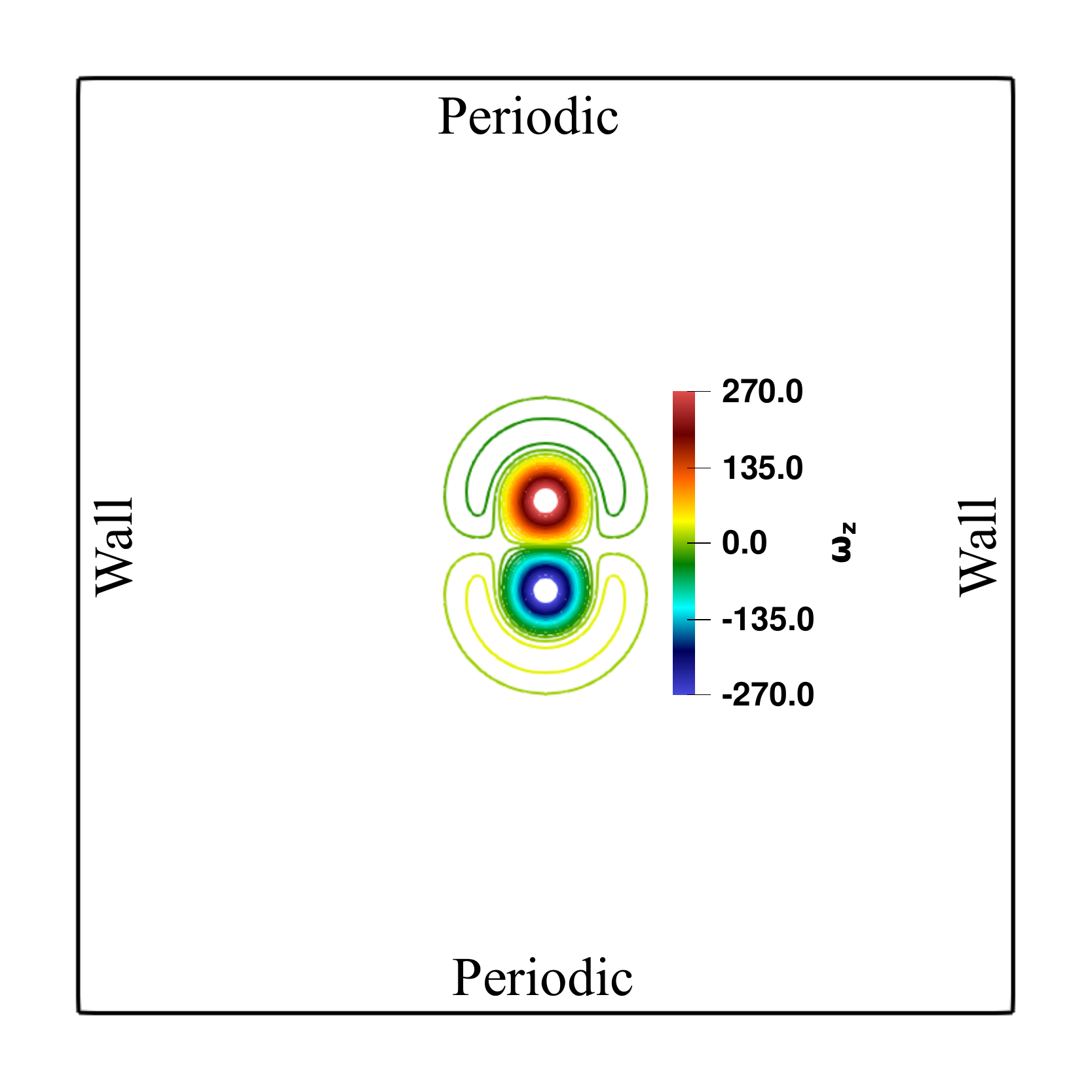}
\caption{The vorticity contours of the initial condition for the dipole-wall collision problem along with the boundary conditions.}
\label{dipole_IC}
\end{figure}

The problem consists of a two-dimensional square domain of size [$2 \times 2$], with the origin at the centre. Figure~\ref{dipole_IC} shows the initial vorticity field along with the boundary conditions. The initial conditions~\cite{keetels2007} are given below,
\begin{subequations}
\begin{align}
    u(x, y, z, t_0) &=  -\frac{1}{2} |\omega_{\max}| (y -y_1) e^{-\left(\frac{r_1}{r_0}\right)^2} + \frac{1}{2} |\omega_{\max}| (y -y_2) e^{-\left(\frac{r_2}{r_0}\right)^2}\\
    v(x, y, z, t_0) &=  \frac{1}{2} |\omega_{\max}| (x -x_1) e^{-\left(\frac{r_1}{r_0}\right)^2} - \frac{1}{2} |\omega_{\max}| (x -x_2) e^{-\left(\frac{r_2}{r_0}\right)^2}\\
    p(x, y, z, t_0) &= 0
\end{align}
\end{subequations}
where, $r_1^2 = (x-x_1)^2 + (y-y_1)^2$ and $r_2^2 = (x-x_2)^2 + (y-y_2)^2$; the extremum vorticity value,  $\omega_{\max} = 299.528385375226$ and radius of the monopoles, $r_0 = 0.1$. The positions of the initial monopoles are given by, ${(x_1,y_1),(x_2,y_2)} = {(0,0.1),(0,-0.1)}$.  The flow is simulated till $t=1.2$ at $Re=1000$. The computational domain is represented with a uniform mesh of $2048^2$.
        
\begin{figure}[H]
    \centering
    \subfigure[]
    {
        \includegraphics[trim = 0mm 5mm 0mm 5mm, clip, width=17cm]{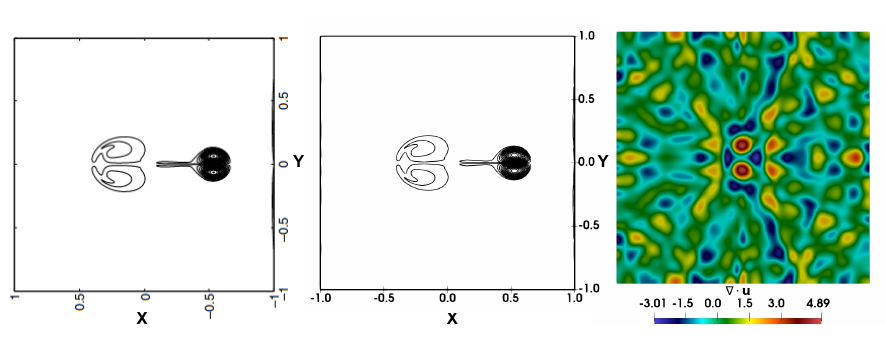}
        \label{fig:vortex_dipole_comp_t0.2}
    }
    \subfigure[]
    {
        \includegraphics[trim = 0mm 5mm 0mm 5mm, clip, width=17cm]{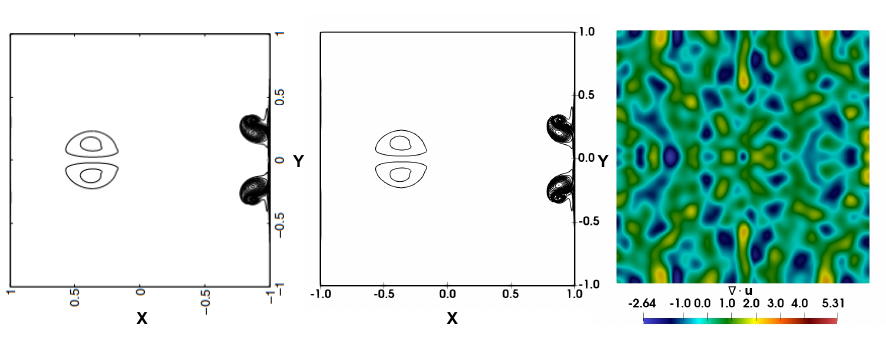}
        \label{fig:Vortex_dipole_comp_t0.4}
    }
    \subfigure[]
    {
        \includegraphics[trim = 0mm 5mm 0mm 5mm, clip, width=17cm]{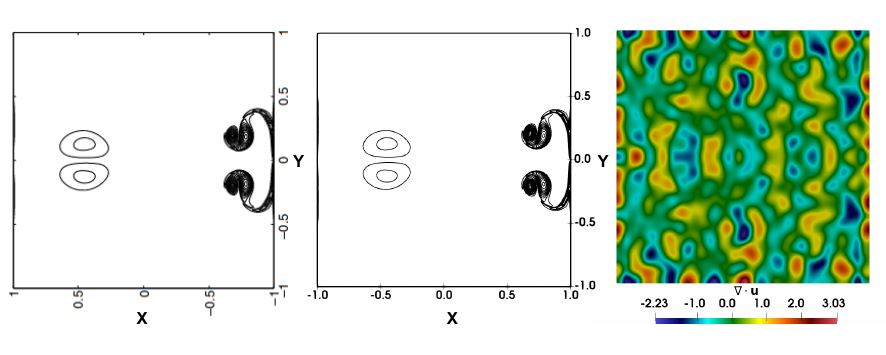}
        \label{fig:Vortex_dipole_comp_t0.5}
    }
\end{figure}
\begin{figure}[H]
    \subfigure[]
    {
        \includegraphics[trim = 0mm 5mm 0mm 5mm, clip, width=17cm]{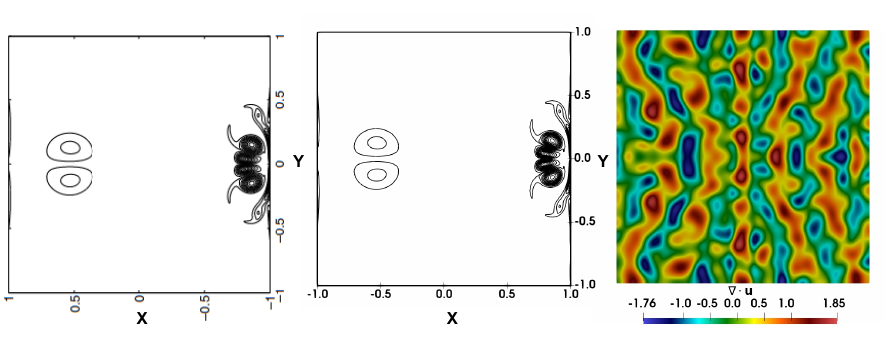}
        \label{fig:Vortex_dipole_comp_t0.6}
    }
    \subfigure[]
    {
        \includegraphics[trim = 0mm 5mm 0mm 5mm, clip, width=17cm]{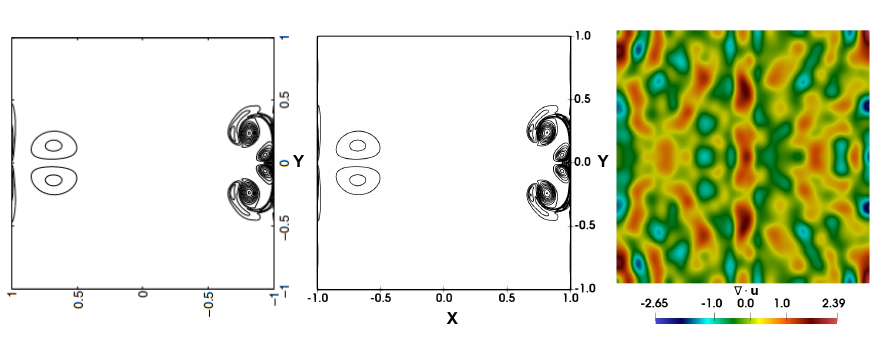}
        \label{fig:Vortex_dipole_comp_t0.8}
    }
    \subfigure[]
    {
        \includegraphics[trim = 0mm 5mm 0mm 5mm, clip, width=17cm]{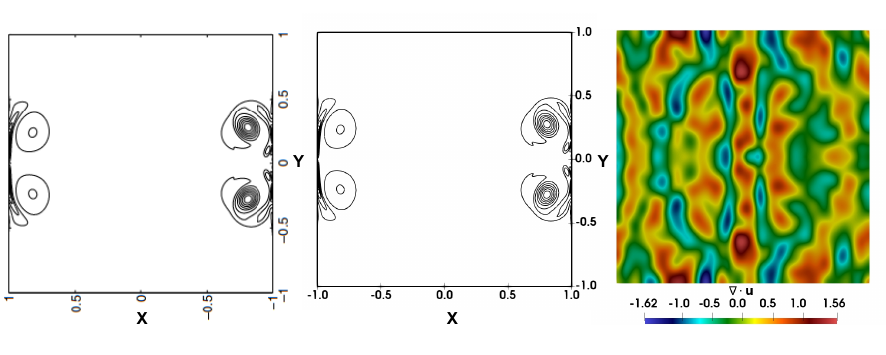}
        \label{fig:Vortex_dipole_comp_t1.2}
    }
    \caption{Contours of vorticity (left \& middle) and velocity divergence (right) for the dipole-wall collision case at  (a) $t=0.2$, (b) $t=0.4$, (c) $t=0.5$, (d) $t=0.6$, (e) $t=0.8$ and (f) $t=1.2$, from Keetels et al.~\cite{keetels2007} (left) and GPE with $Ma=0.02$ (middle).  The contour levels are drawn for $\boldsymbol{\omega_z} = \pm 10$, $\pm 30$, $\pm 50$, $\pm 70$, $\pm 90$, $\pm 110$, $\pm 130$, $\pm 150$, $\pm 170$, $\pm 190$, $\pm 210$, $\pm 230$, $\pm 250$ and $\pm 270 $. Results from Keetels et al.~\cite{keetels2007} are reproduced with permission from the publisher.}
    \label{dipoleWall_vortContour}
\end{figure}

As the flow evolves, the dipole hits the right wall and rebounds. As mentioned earlier, it is challenging to predict the post-collision coherent vortical structures. First, we qualitatively examine the performance of the present weakly compressible approach with $Ma=0.02$ by a rigorous comparison of the vorticity field's evolution with the results published in the literature. The vorticity contours at various time instances are illustrated in figure~\ref{dipoleWall_vortContour}, together with those reported by Keetels et al.~\cite{keetels2007}, who used high-accurate Chebyshev spectral methods. At $t=0.2$, the dipoles move towards the wall and as a result of the subsequent collision, a new set of secondary vortices is generated, as seen in figure~\ref{fig:Vortex_dipole_comp_t0.4}. These new vortices again collide with the wall at around $t=0.6$. After $t=0.8$, no more collision happens, and the existing vortical structures slowly decay. As mentioned earlier, in spite of the complicated flow dynamics, we observe an excellent match between these contour plots at all the time instants.

\begin{figure}[!ht]
\subfigure[]
{
    \includegraphics[trim = 0mm 0mm 0mm 0mm, clip, width=17cm]{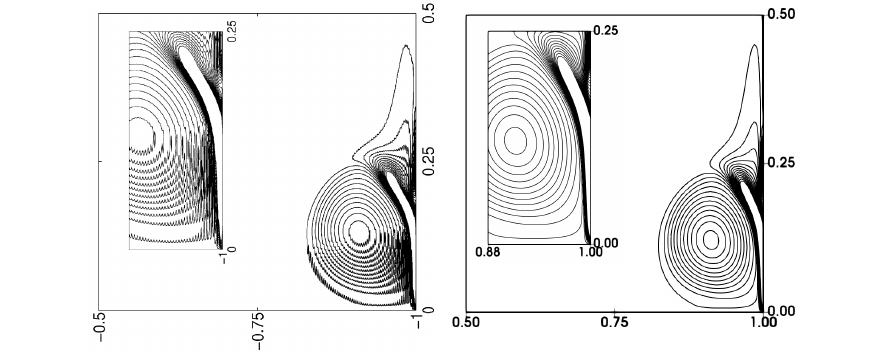}
    \label{fig:Vortex_dipole_comp_t0.35_Zoom}
}
\subfigure[]
{
    \includegraphics[trim = 0mm 0mm 0mm 0mm, clip, width=17cm]{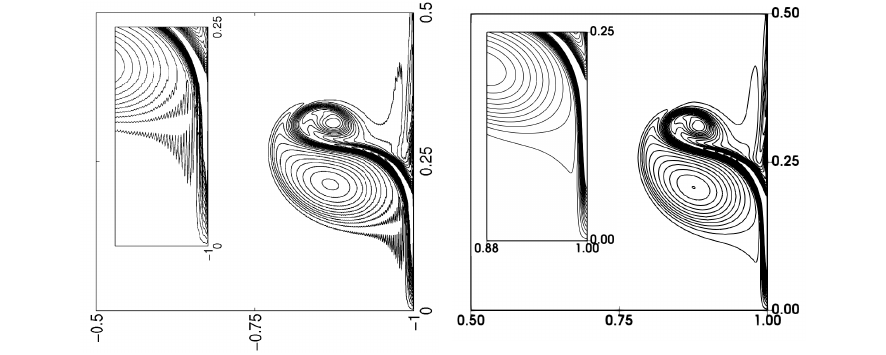}
    \label{fig:Vortex_dipole_comp_t0.4_Zoom}
}
\caption{Close-up view of vorticity contours for the dipole-wall collision test case at (a) $t=0.35$, (b) $t=0.4$ from Keetels et al.~\cite{keetels2007} (left) and GPE with $Ma=0.02$ (right). Results from Keetels et al.~\cite{keetels2007} are reproduced with permission from the publisher.}
\label{dipoleWall_vortContour_Zoom}
\end{figure}

A crucial point of observation is the ``unacceptably" large values of the mass conservation error for the present example, as can be inferred from the plots of $\nabla \cdot \textbf{u}$ presented in figure~\ref{dipoleWall_vortContour}. For all the instances reported, $|\nabla \cdot \textbf{u}|>1.5$, and it reaches as large as 5.31 as seen in figure~\ref{fig:Vortex_dipole_comp_t0.4}. Despite such a large error, the agreement with the reported results is excellent. To provide an even more careful comparison, we present, in figure~\ref{dipoleWall_vortContour_Zoom}, an enlarged view of the vorticity contours and compare it with those reported by Keetels et al.~\cite{keetels2007}. This provides further evidence that the results from GPE are quite accurate.

These results are surprising and intriguing.  In general, such a large error in mass conservation is considered unacceptable in the CFD community; it is expected that $\nabla \cdot \textbf{u}$ is at least $10^{-3}$ or smaller. Hence, with such large errors, one might expect either a complete breakdown of the simulation or a divergence of the numerical solution from the physical scenario leading to unphysical results.  Nevertheless, we find that the higher mass conservation error has no visible effect on the flow field, even for such an extremely challenging test case. Hence, our results prove that weakly compressible approaches can yield physically correct flowfields, although they do not accurately satisfy mass conservation.

It is to be noted that for the DPSL problem discussed in the previous section, $\nabla \cdot \textbf{u}\sim 0.1$ for $Ma=0.2$, and this leads to an inaccurate solution. Here, although, $\nabla \cdot \textbf{u}\sim 5$ for $Ma=0.02$, accurate results are obtained. This confirms that the mass conservation error alone does not dictate the accuracy of the weakly compressible methods.

\begin{figure}[!ht]
\centering
\subfigure[]
{
    \includegraphics[trim = 0mm 0mm 0mm 0mm, clip, width=7.5cm]{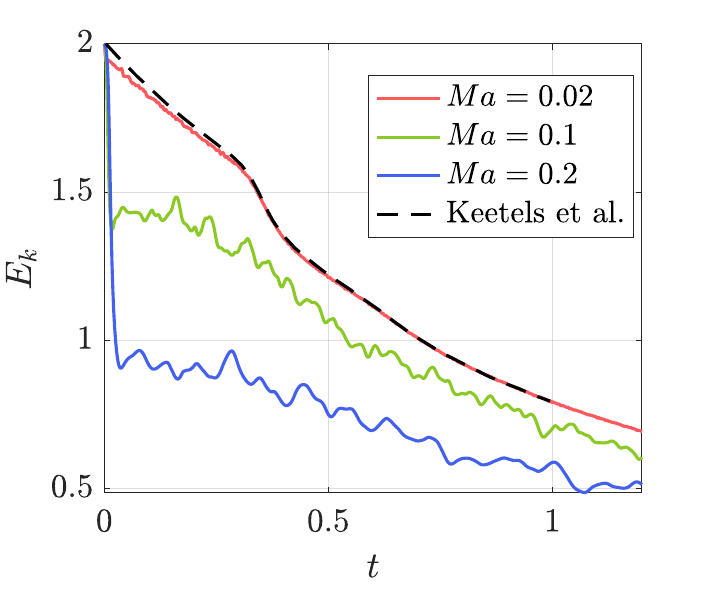}
}
\subfigure[]
{
    \includegraphics[trim = 0mm 0mm 0mm 0mm, clip, width=7.5cm]{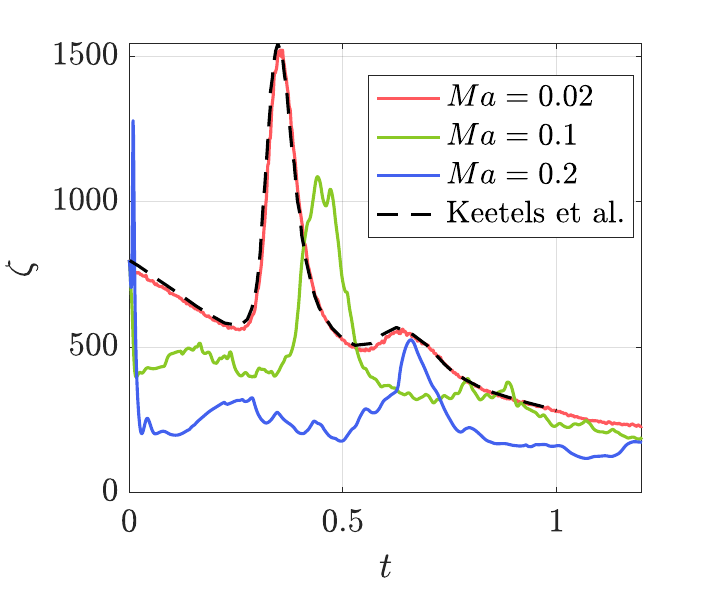}
}
\caption{Evolution of (a)~kinetic energy ($E_k$) and (b)~enstrophy ($\zeta$) for the dipole-wall collision test case from GPE compared with those from Keetels et al.~\cite{keetels2007}.}
\label{dipoleWall_Ke_en_512}
\end{figure}

Until now, we have provided a qualitative comparison of the results obtained from GPE with that reported in the literature. For a quantitative analysis, we investigate the evolution of kinetic energy and enstrophy. These plots are presented in  figure~\ref{dipoleWall_Ke_en_512} for all $Ma$ considered. As expected from the qualitative comparison presented earlier, results from $Ma=0.02$ provides excellent agreement for both $E_k$ and $\zeta$ with the reference data from Keetels et al.~\cite{keetels2007}. However, as $Ma$ increases, there appears a sharp dip in $E_k$ and a large peak in $\zeta$ at the very beginning of the simulation. This observation affirms the presence of a strong (non-physical) dissipation mechanism, which will be discussed next.

\begin{figure}[!ht]
\centering
\subfigure[]
{
    \includegraphics[trim = 0mm 15mm 0mm 0mm, clip, width=15cm]{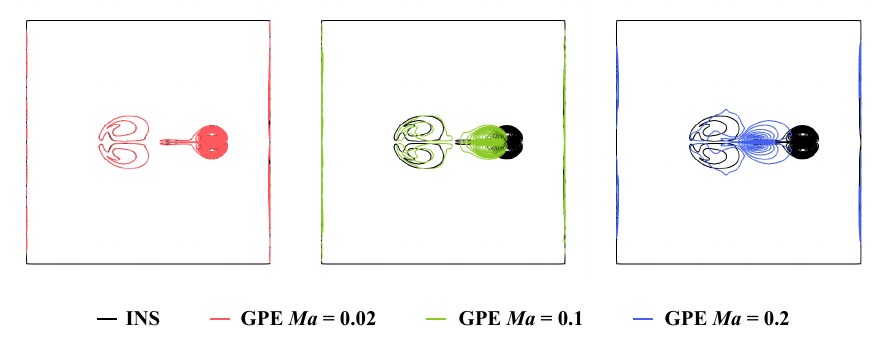}
    \label{dipoleWall_Ma_study_flowField_t0.2}
}
\subfigure[]
{
    \includegraphics[trim = 0mm 0mm 0mm 0mm, clip, width=15cm]{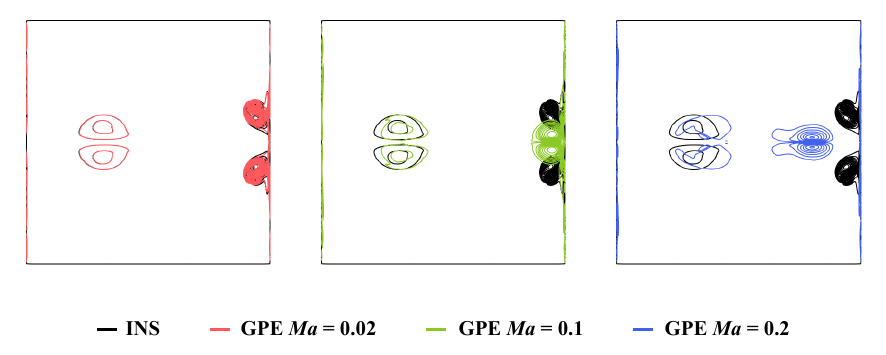}
    \label{dipoleWall_Ma_study_flowField_t0.4}
}
\caption{Contours of vorticity for the dipole-wall collision case at (a) $t=0.2$ and (b) $t=0.4$ from the GPE solver for different values of Mach number. `INS' indicates incompressible flow results obtained using OpenFOAM. The contour levels are the same as that of Figure~\ref{dipoleWall_vortContour}. }
\label{dipoleWall_Ma_study_flowField}
\end{figure}

In order to visually illustrate the presence of a strong spurious dissipation, we simulated this problem using the popular open-source tool OpenFOAM v2012. We compare the vorticity contours of incompressible flow results obtained using OpenFOAM and the weakly compressible results obtained from GPE with different $Ma$. These contours are presented in figure~\ref{dipoleWall_Ma_study_flowField} for $t=2$ and $t=4$, which correspond to instances before and after the collision, respectively, for $Ma=0.02$. The results from $Ma=0.02$ are in excellent agreement with that of the incompressible flow. However, as $Ma$ increases, the movement of the vortex-dipole towards the wall gets slower, which is clearly evident for both $Ma=0.1$ and 0.2 as observed from figure~\ref{dipoleWall_Ma_study_flowField_t0.2}. At $t=4$, we observe that the dipole in the $Ma=0.1$ case is seen to lag behind and is in the process of colliding with the wall. At both the time instances, for $Ma=0.2$, the vortex dipoles are weak and less intense, and even at $t=4$, they are yet to reach the wall.
      
\begin{figure}[!ht]
\centering
\subfigure[]
{
    \includegraphics[trim = 0mm 0mm 0mm 0mm, clip, width=7.5cm]{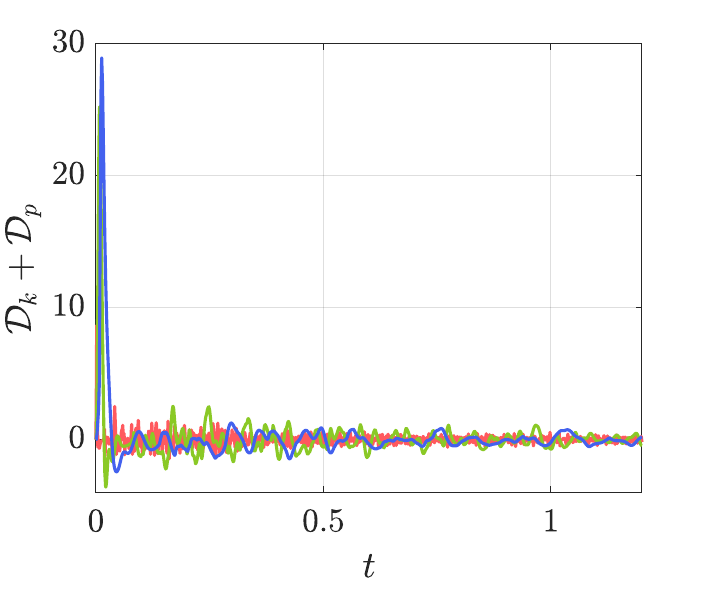}
}
\subfigure[]
{
    \includegraphics[trim = 0mm 0mm 0mm 0mm, clip, width=7.5cm]{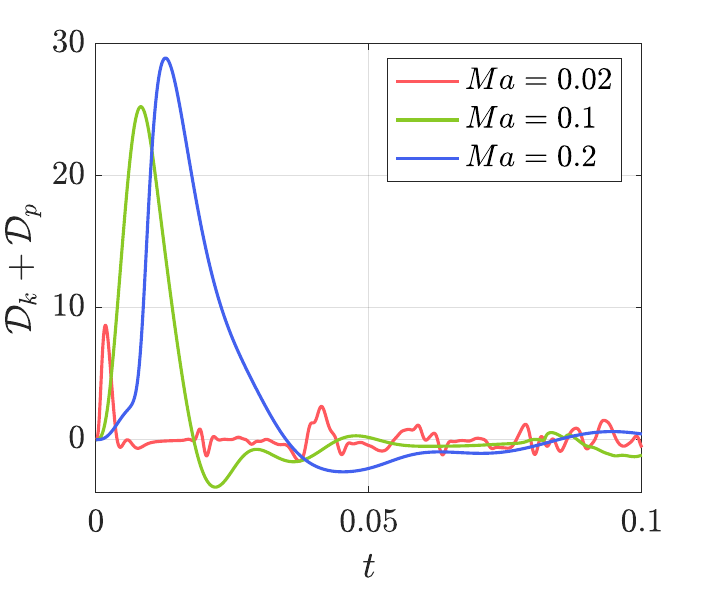}
}
\caption{The sum of dilatation terms for the dipole-wall collision problem: (a)~for the entire duration of the simulation, and (b)~zoomed view until $t=0.1$.}
\label{fig:dila_dipole}
\end{figure}

The aforementioned flow field details clearly confirm the existence of spurious dissipation at the start of the simulation. To delineate the dissipation mechanism, we compute the sum of the dilatation terms~($\mathcal{D}_k+\mathcal{D}_p$), and depict its temporal evolution in figure~\ref{fig:dila_dipole}. It shows a distinct peak at the start of the simulation for all $Ma$, consistent with $E_k$ and $\zeta$~(figure~\ref{dipoleWall_Ke_en_512}). For brevity, a close-up view of the plot until $t=0.1$ is presented in figure~\ref{fig:dila_dipole}(b). Two important observations can be made: (i)~the peak value is more than two orders of magnitude larger than DPSL, and (ii)~ the time scale over which the peak occurs is much smaller than DPSL, the consequence of which we discuss next.

\begin{figure}[!ht]
\begin{center}
\includegraphics[trim = 0mm 0mm 0mm 0mm, clip, width=15cm]{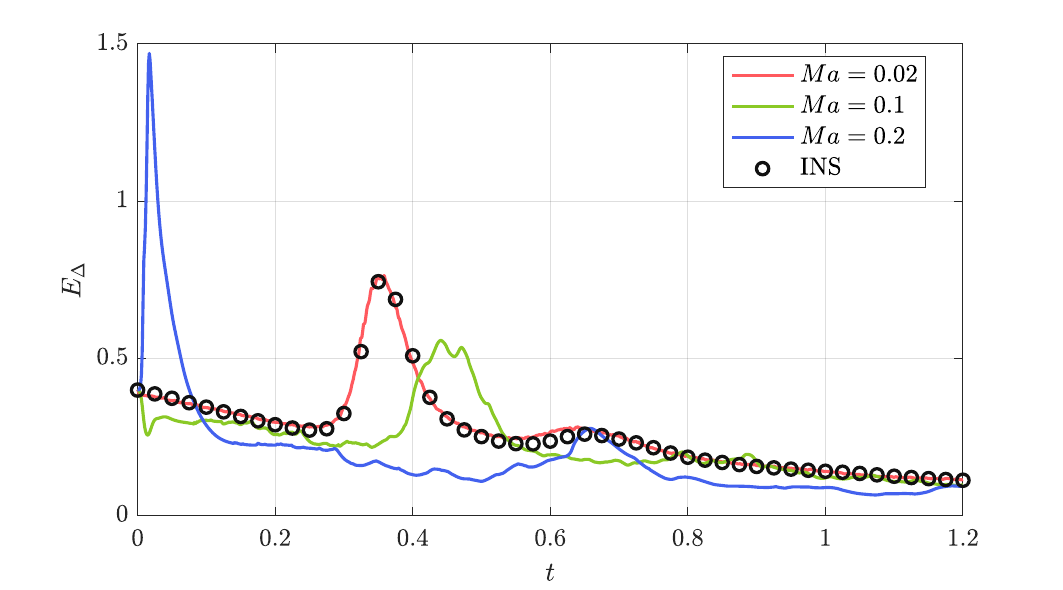}
\end{center} 
\caption{Rate of viscous dissipation for the dipole-wall collision problem. The plot denoted as `INS' represents incompressible OpenFOAM simulation.}
\label{dipole_dissi}
\end{figure}

We present the variation of the rate of viscous dissipation for all $Ma$ in figure~\ref{dipole_dissi}. Here also, the $Ma=0.02$ results match well with INS, while the other Mach numbers show different behaviour. In both the $\zeta-$ plot~(figure~\ref{dipoleWall_Ke_en_512}(b)) and $E_\Delta-$ plot~(figure~\ref{dipole_dissi}), we clearly observe that the physical peak is diminished for $Ma=0.1$ and 0.2, implying the strength of the dipole is weakened; moreover, the time lag in the peak indicates that the time at which the dipole impinges the wall is delayed.

To quantify the effect of the dilatation-induced dissipation, we trace the time~($\tilde{t}$) at which the $\mathcal{D}_k+\mathcal{D}_p$ curve crosses zero after the first peak. We evaluate the total contribution of the peak by defining an integral quantity, for a general variable $\mathcal{X}$ as follows
\begin{equation}
    \widetilde{\mathcal{X}}=\int_0^{\tilde{t}}{\mathcal{X} dt}
\end{equation}
The integral value of the total dissipation rate from GPE, and the viscous rate of dissipation obtained from incompressible flow results, denoted as INS, are given in table~\ref{tab:dipole_dissi}. The total dissipation rate, even for $Ma=0.02$, is more than ten times that of INS. However, since the time scale over which the peak occurs is very small, it yielded only a minor dip in $E_k$~(figure~\ref{dipoleWall_Ke_en_512}). Eventually, as time progresses, the results follow that of incompressible flow. For both $Ma=0.1$ and 0.2, the total dissipation rate is approximately 30 times that of the INS value. Since the timescale for $Ma=0.2$ is larger than 0.1, it experiences a larger dip in $E_k$.

\begin{table}[!ht]
\centering
\caption{The integral value of the total dissipation rate from GPE at various $Ma$, and the viscous dissipation rate obtained from incompressible flow results.}
\footnotesize
\begin{tabular}{cccc}
    \hline
    \hline\\
    $Ma$ & $\tilde{t}$  & $\widetilde{\mathcal{D}_k}+\widetilde{\mathcal{D}_p}+\widetilde{E_\Delta}$ & 
    $\widetilde{E_\Delta}$ from INS\\ \hline 
0.02 & 0.0039 & 0.0183 & 0.0015  \\
0.1  & 0.0179 & 0.2283 & 0.0071 \\
0.2  & 0.0348 & 0.3963 & 0.0136 \\
    \hline
    \hline
\end{tabular}
\label{tab:dipole_dissi}
\end{table}

\begin{figure}[!ht]
\centering
    \includegraphics[trim = 0mm 0mm 0mm 0mm, clip, width=15cm]{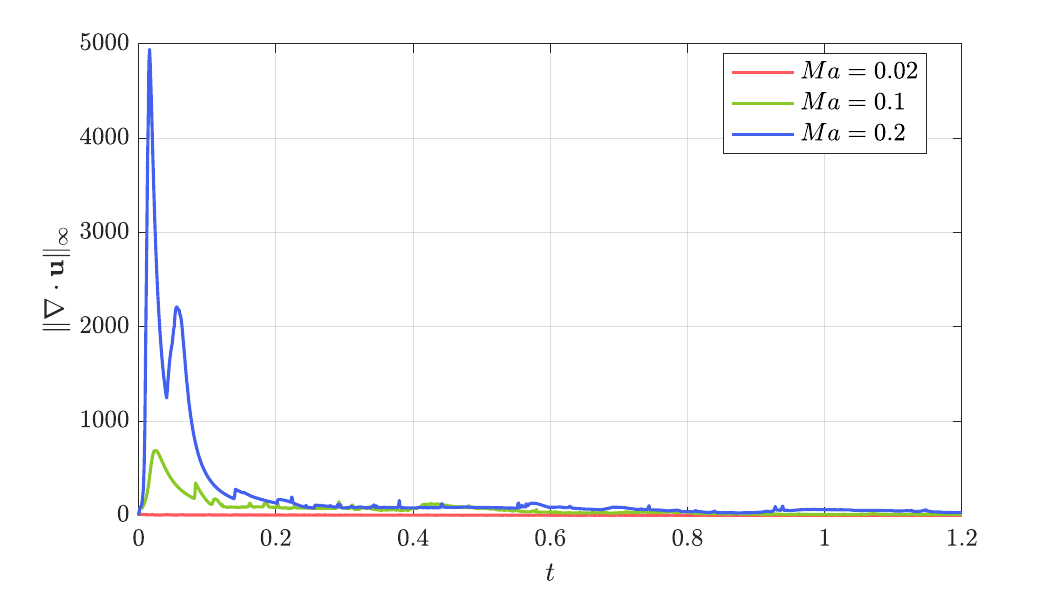}
\caption{Evolution of $L_{\infty}$ norm of velocity divergence for the dipole-wall collision problem from the GPE solver.}
\label{dipoleWall_velDiv}
\end{figure}
\begin{figure}[!ht]
\centering
\subfigure[]
{
    \includegraphics[trim = 0mm 0mm 0mm 0mm, clip, width=5cm]{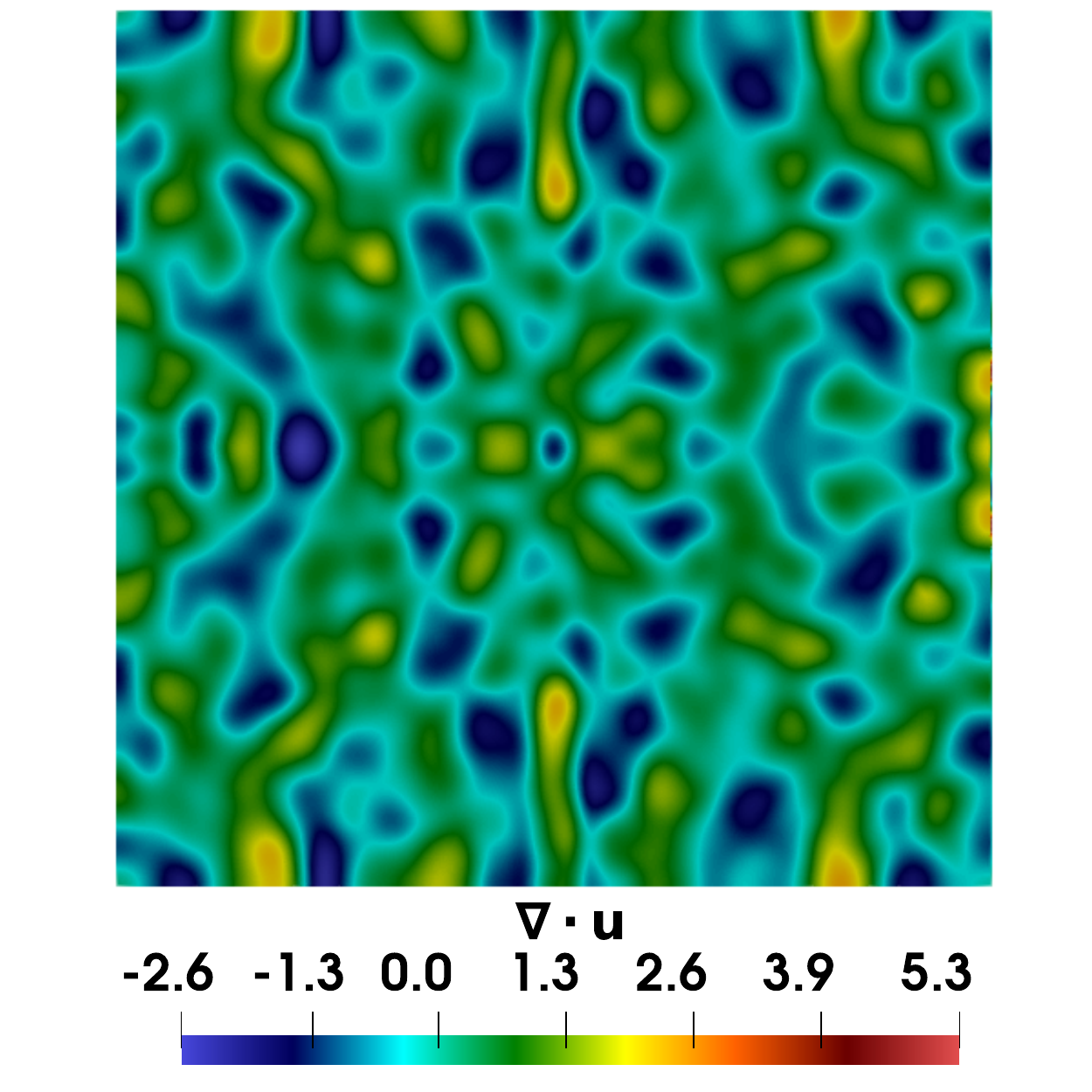}
}
\subfigure[]
{
    \includegraphics[trim = 0mm 0mm 0mm 0mm, clip, width=5cm]{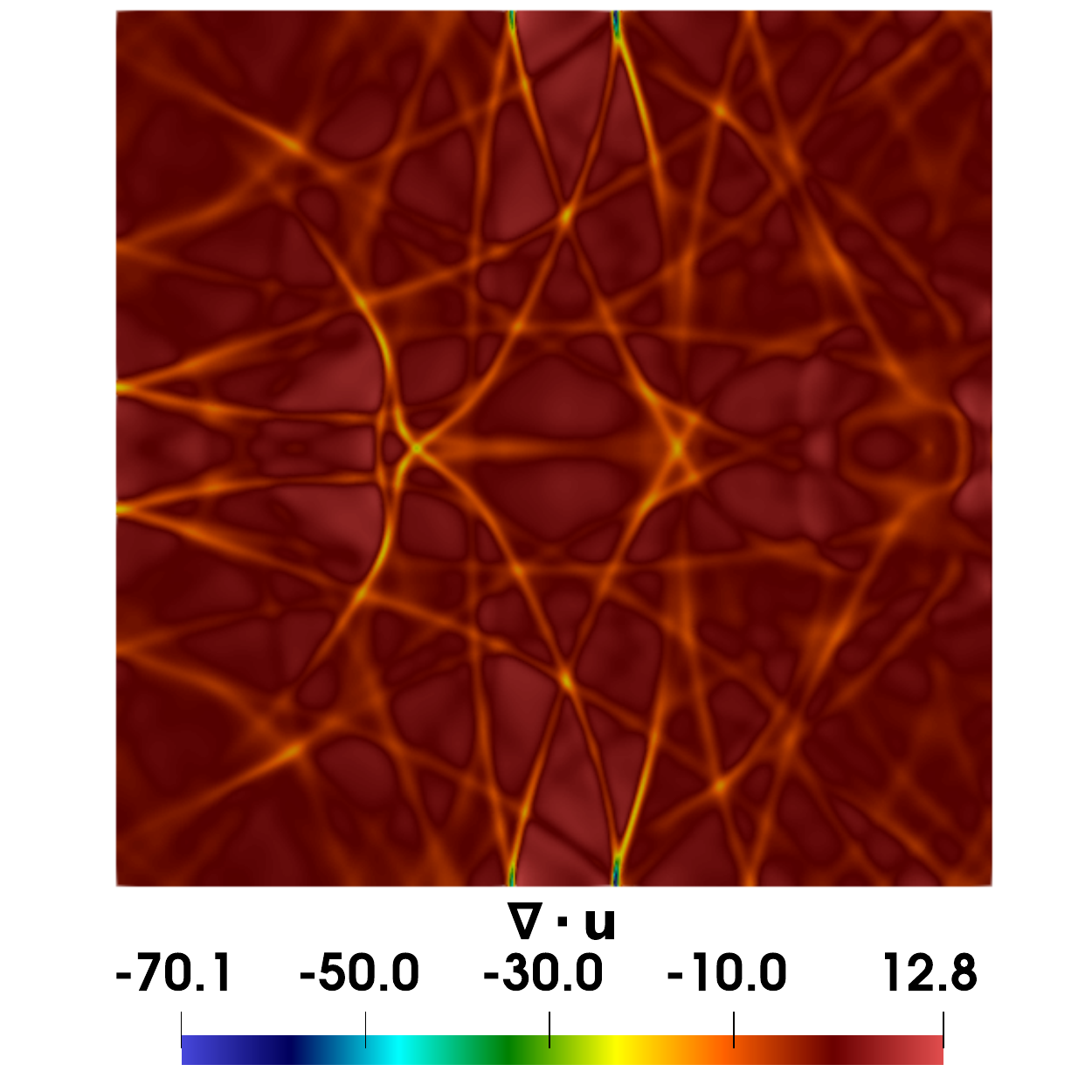}
}
\subfigure[]
{
    \includegraphics[trim = 0mm 0mm 0mm 0mm, clip, width=5cm]{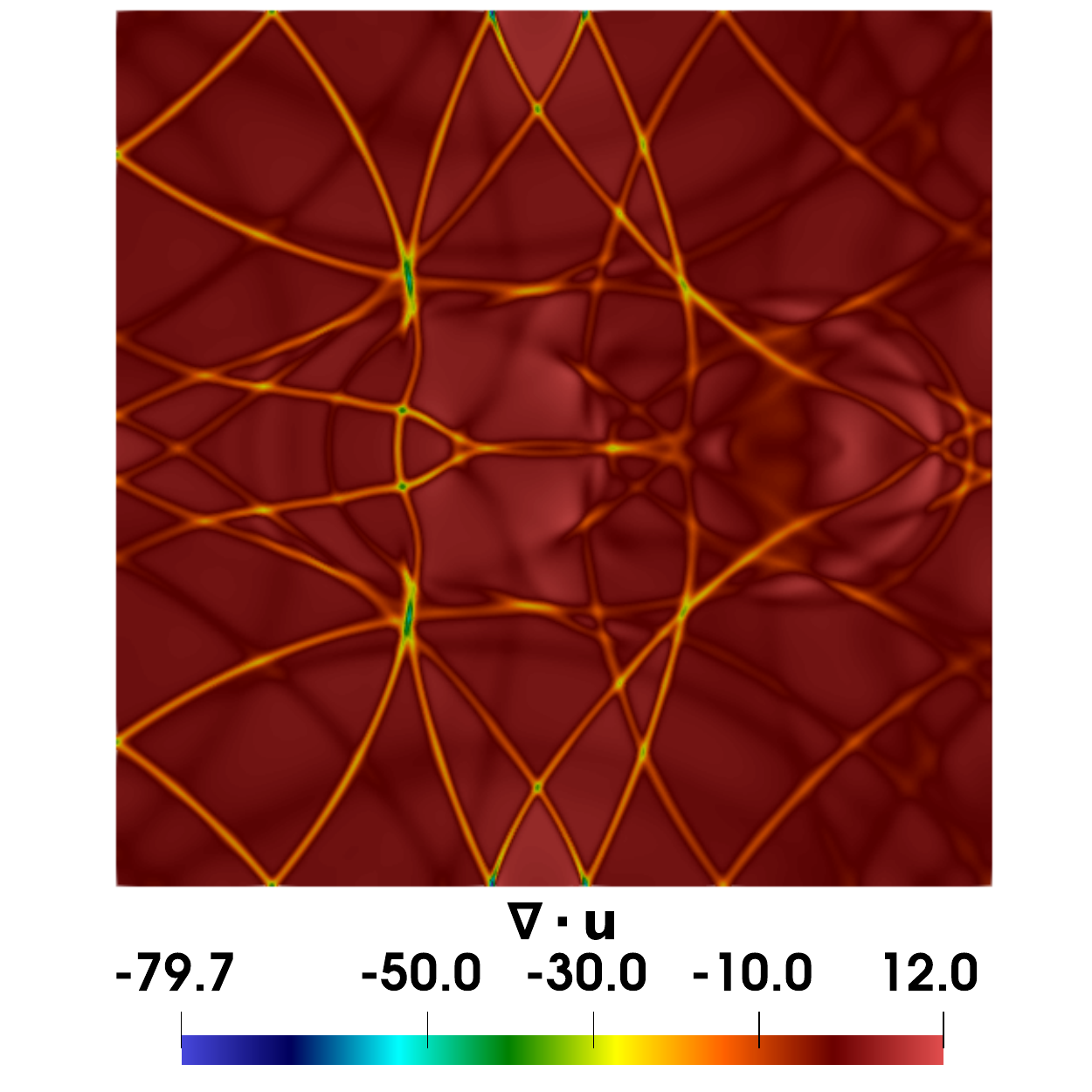}
}
\caption{Palettes of $\nabla\cdot\textbf{u}$ for the dipole-wall collision problem at $t=0.4$: (a)~$Ma=0.02$, (b)~$Ma=0.1$, and (c)~$Ma=0.2$.}
\label{fig:veldiv_dipole_t0.4}
\end{figure}

We noted earlier that for $Ma=0.02$, the mass conservation error reaches more than 5. It is of interest to know the values of $\nabla\cdot\textbf{u}$ for the other $Ma$. From the temporal evolution plot of the $L_\infty-$norm of $\nabla\cdot\textbf{u}$, presented in figure~\ref{dipoleWall_velDiv}, we see that for $Ma=0.2$, it reaches close to 5000. Despite such a huge error, the simulation did not have any convergence issue, because this error, through the dilatation terms, adds dissipation to the flow field that ``stabilizes'' the flow. With such a large dilatation, we expect pronounced acoustic pressure waves in the flow field. This is evident from the palettes of $\nabla\cdot\textbf{u}$, presented in figure~\ref{fig:veldiv_dipole_t0.4}, at $t=0.4$. The acoustic waves observed for $Ma=0.1$ and 0.2 are significantly stronger. Although for $Ma=0.02$, the waves are weaker, they induce large oscillations in the forces acting on a solid immersed in the fluid. This has been addressed in detail in the recent work by Raghunathan and Sudhakar~\cite{raghunathan2024}.
	
The weakly compressible methods are extensively validated with numerical examples ranging from laminar~\cite{Toutant2018, Clausen2013}, turbulent~\cite{Dupuy2020,Shi2020,kajzer2018}, moving and deforming domains~\cite{bolduc2023}, two-phase~\cite{bodhanwalla2024, kajzer2022, yang2021}, buoyancy-driven flows~\cite{sharma2023}, etc. Despite the high popularity of these methods, a test case with a significant mass conservation error is yet to be reported. In this study, we purposefully chose the dipole wall collision problem to illustrate a scenario with a notably high mass conservation error and emphasize that the flow structures remain unaffected despite such large errors, provided the dilatation dissipation's contribution is not significant.

\subsection{Turbulent Taylor Green vortex}
The previous test cases reported the influence of the dilatation terms in laminar flows. In order to extend our investigation to turbulent flows, we simulate the three-dimensional unsteady turbulent Taylor-Green vortex (tTGV) problem. This numerical example serves the following two purposes: (i)~investigation of dilatation-driven dissipation for a turbulent flow, and (ii)~analysis of the accuracy of GPE-based weakly compressible approaches to model unsteady turbulent flows.

The flow, in this test case, is confined to a periodic cubic domain of size [$2\pi \times 2\pi \times 2\pi$]. Following Bull et al.~\cite{bull2015}, the initial velocity  and pressure fields are given by
\begin{subequations}
\begin{align}
u(x, y, z, t_0) &= \: \: V_0 \: \sin \left( \frac{x}{L_0} \right) \cos \left( \frac{y}{L_0} \right) \cos \left( \frac{z}{L_0} \right), \\
v(x, y, z, t_0) &= -V_0 \: \cos \left( \frac{x}{L_0} \right) \sin \left( \frac{y}{L_0} \right) \cos \left( \frac{z}{L_0} \right), \\
w(x, y, z, t_0) &= \: \: 0, \\
p(x, y, z, t_0) &= \: \: p_0 + \frac{\rho_0}{16}\left[ \cos \left( \frac{2x}{L_0}\right) + \cos \left( \frac{2y}{L_0}\right) \right] \left[ \cos \left( \frac{2z}{L_0} \right) + 2 \right].
\end{align}
\end{subequations}
Here, $L_0$, $V_0$, $p_0$ and $\rho_0$ are taken as unity. We set these initial conditions and allow the flow to evolve. We simulate the flow till $t = 20$ at $Re=1600$.

As the first objective, we investigate the influence of the dilatation-induced dissipation terms on a coarse $256^3$ mesh. To compare the influence of these spurious terms relative to the physically correct rate of viscous dissipation, we require results of DualACM that are prohibitively expensive on a much-refined mesh for this three-dimensional problem. Hence, we limit our discussion to a coarse mesh for this objective.

\begin{figure}[H]
    \centering
        \includegraphics[trim = 0mm 0mm 0mm 0mm, clip, width=15cm]{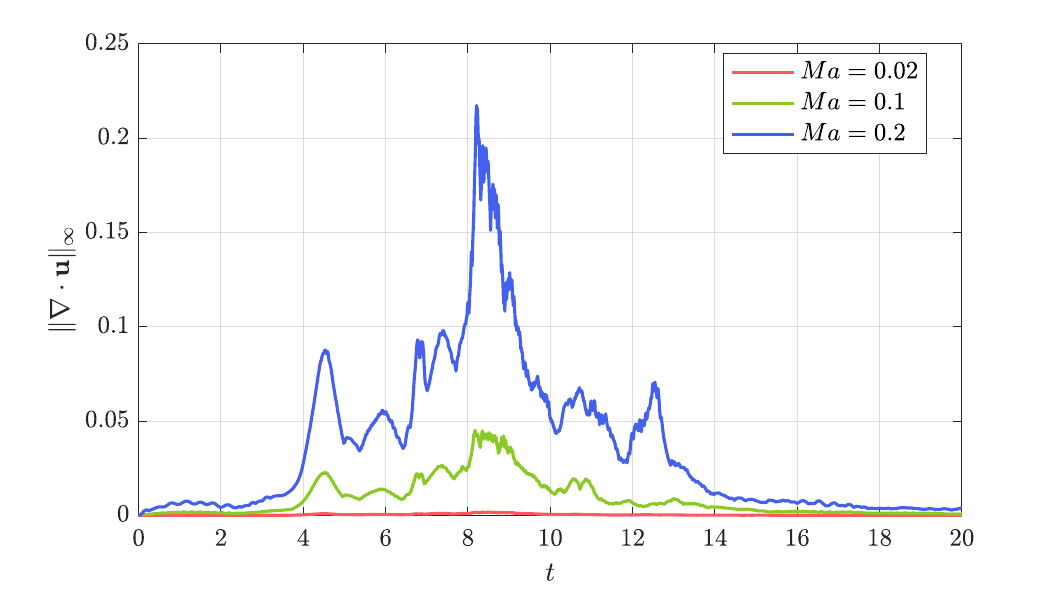}
        \label{tTGV_velDivMax_256}
    \caption{Evolution of mass conservation error for the turbulent Taylor-Green vortex problem using $256^3$ mesh. The error is represented in terms of $L_\infty-$norm.}
    \label{tTGV_velDiv_256}
\end{figure} 

\begin{figure}[!ht]
		\centering
		\subfigure[]
		{
			\includegraphics[trim = 0mm 0mm 0mm 0mm, clip, width=7.5cm]{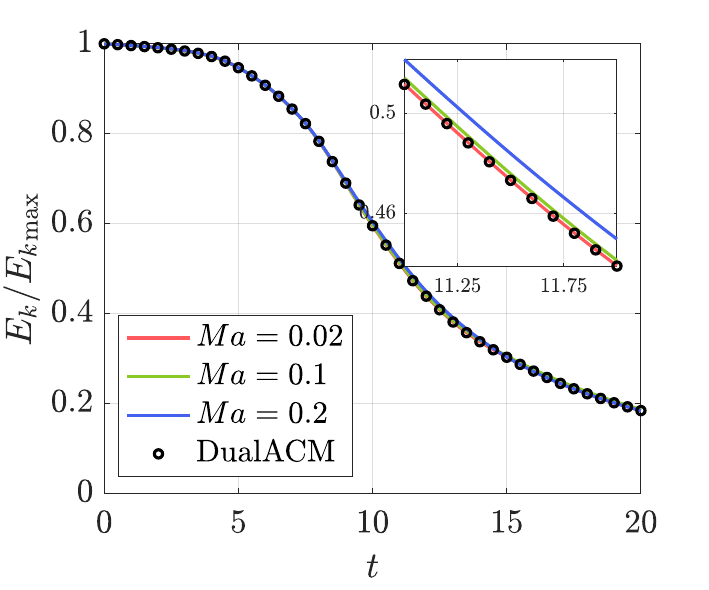}
		}
		\subfigure[]
		{
			\includegraphics[trim = 0mm 0mm 0mm 0mm, clip, width=7.5cm]{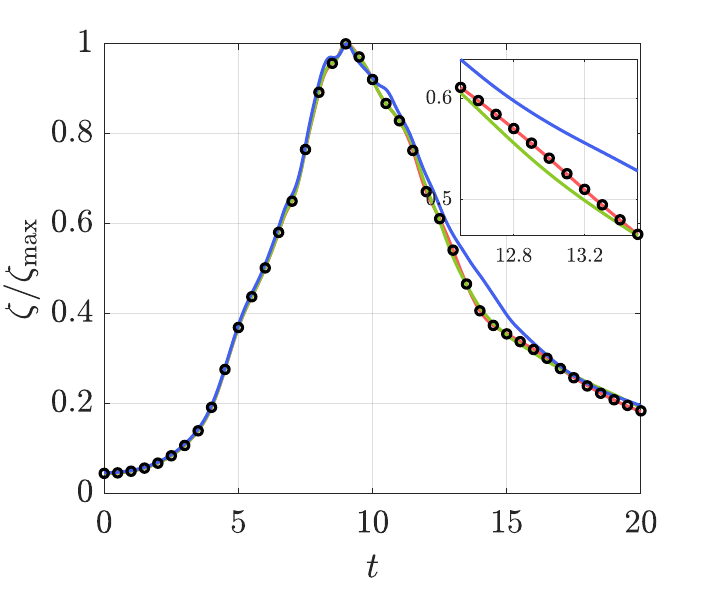}
		}
		\caption{Evolution of (a)~kinetic energy~($E_k$) and (b)~enstrophy~($\zeta$) obtained from GPE and DualACM~($\epsilon=10^{-6}$) schemes for the turbulent Taylor-Green vortex problem using $256^3$ grid.}
		\label{tTGV_enstrophy256}
	\end{figure}

The mass conservation error, in terms of $L_\infty-$norm, for all $Ma$ are presented in figure~\ref{tTGV_velDiv_256}. Similar to the findings of AbdulGafoor et al.~\cite{abdulgafoor2024} for EDAC, we also observe a direct relation between the velocity divergence and the associated flow physics. During the laminar regime ($t<5$), the velocity divergence slowly increases, followed by a rapid rise during the transition period ($5<t<9$), and it decreases as the turbulent structures decay ($9<t<20$). Similar to AbdulGafoor et al.~\cite{abdulgafoor2024}, we observe a peak in velocity divergence when the enstrophy, presented in figure~\ref{tTGV_enstrophy256} is the highest. The velocity divergence, $\nabla\cdot\textbf{u}\sim 10^{-3}$ for $Ma=0.02$, hence it is not visible in figure~\ref{tTGV_velDiv_256}.  An important observation is that the maximum value of $\nabla\cdot\textbf{u}$ reaches more than 0.2 for $Ma=0.2$, and is in the range of the DPSL problem discussed in section~\ref{sec:dpsl}. Hence, similar to that test case, one might expect a significantly dissipated solution for $Ma=0.2$.

\begin{figure}[!ht]
\centering
\subfigure[]
{
    \includegraphics[trim = 0mm 0mm 0mm 0mm, clip, width=7.5cm]{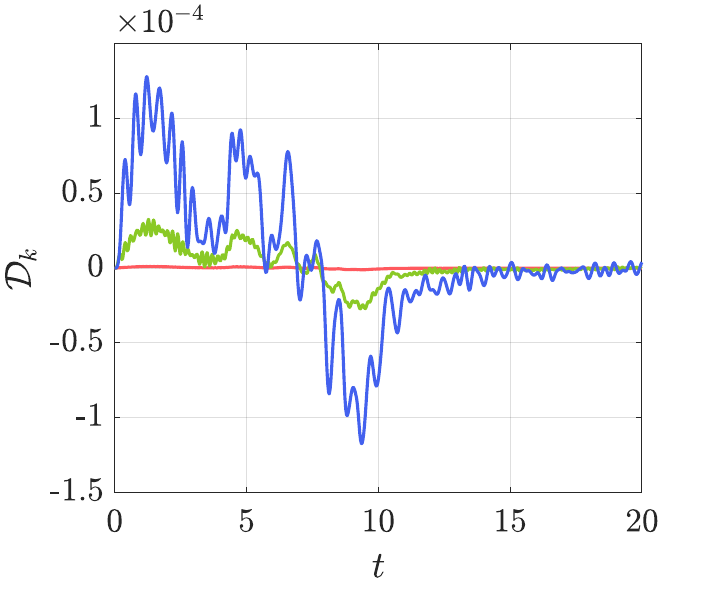}
}
\subfigure[]
{
    \includegraphics[trim = 0mm 0mm 0mm 0mm, clip, width=7.5cm]{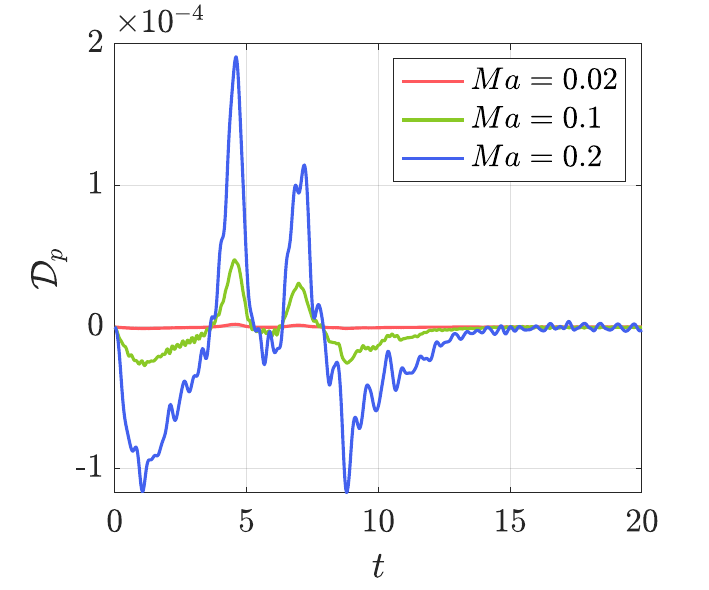}
}
\caption{Evolution of (a)~kinetic energy dilatation and (b)~pressure dilatation for the turbulent Taylor-Green vortex on $256^2$ mesh.}
\label{tgv_ke_p_dila}
\end{figure}

The evolution of $\mathcal{D}_k$ and $\mathcal{D}_p$ are presented in figure~\ref{tgv_ke_p_dila}. The magnitude of both terms is of the order of $10^{-4}$ even for $Ma=0.2$. It is clear that both the dilatation terms are much weaker here compared to the previous test cases. This is an important test case because $\nabla\cdot\textbf{u}$ is of the order of DPSL, but $\mathcal{D}_k$ and $\mathcal{D}_p$ are two orders of magnitude smaller. Hence, this example is an ideal candidate to shed light on the importance of these parameters that arise only in weakly compressible methods.

\begin{figure}[!ht]
\begin{center}
\includegraphics[trim = 5mm 0mm 10mm 5mm, clip, width=12cm]{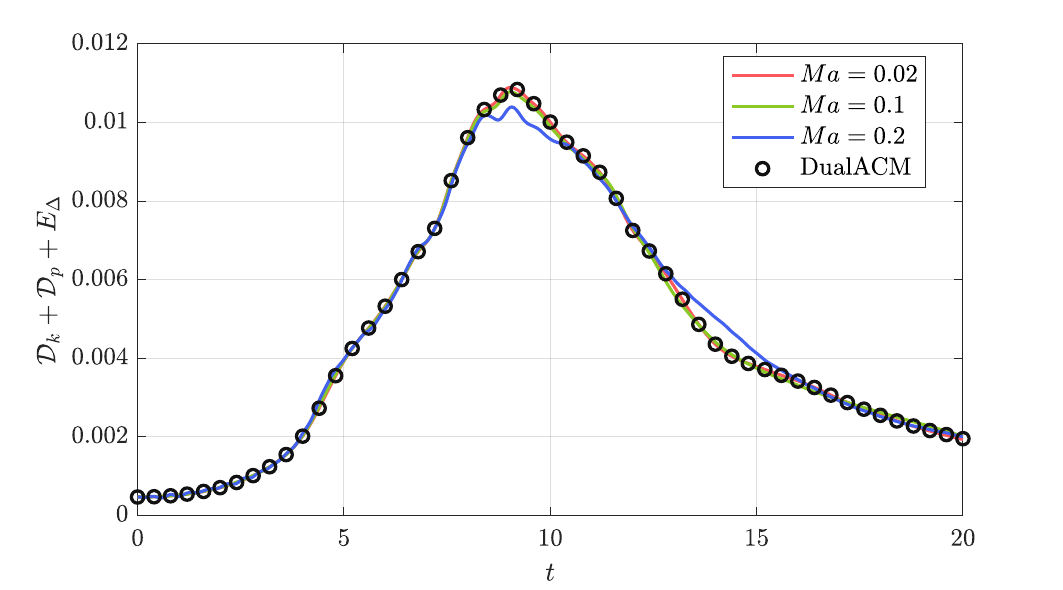}
\end{center} 
\caption{Total kinetic energy dissipate rate per unit volume for the turbulent Taylor-Green vortex on $256^2$ mesh.}
\label{fig:tgv_dissi}
\end{figure} 

Turbulent flows are characterised by their extreme sensitivity to infinitesimal perturbations. Hence, we might expect that even such a small disruption to the physical dissipation can significantly alter the flowfield evolution. To verify this, we trace the total energy dissipation rate and present it in figure~\ref{fig:tgv_dissi}. The results from $Ma=0.02$ and 0.1 match very closely with that of the incompressible flow represented by DualACM. The simulation with $Ma=0.2$ shows only minor variation compared to the previous test cases. The percentage difference in the total dissipation rate is approximately 5.5\% for the peak dissipation observed at $t\approx 9$.

Although the mass conservation errors are of the same orders of magnitude as the doubly periodic shear layer example, since the dilatation-induced dissipation is not significant enough, the obtained results match very well with the reference DualACM results even for $Ma=0.1$ and 0.2. Plots of the evolution of kinetic energy~(figure~\ref{tTGV_enstrophy256}a) and enstrophy~(figure~\ref{tTGV_enstrophy256}b) reinforce this interpretation. We can draw a crucial conclusion from these observations: it is the dilatation-driven dissipation terms, not the mass conservation error alone, dictate the accuracy of WCMs.

While the exact reason for the diminished magnitude of $\mathcal{D}_k$ and $\mathcal{D}_p$ is not clearly known, it is consistent with the results reported by Liu et al.~\cite{liu2016}, who solved the compressible Navier-Stokes equations using the discontinuous Galerkin method. Their results with $Ma=0.1$ also show a non-significant contribution of the pressure dilatation term to the total dissipation.

	\begin{figure}[!ht]
		\centering
		\subfigure[]
		{
			\includegraphics[trim = 20mm 0mm 25mm 0mm, clip, width=7cm]{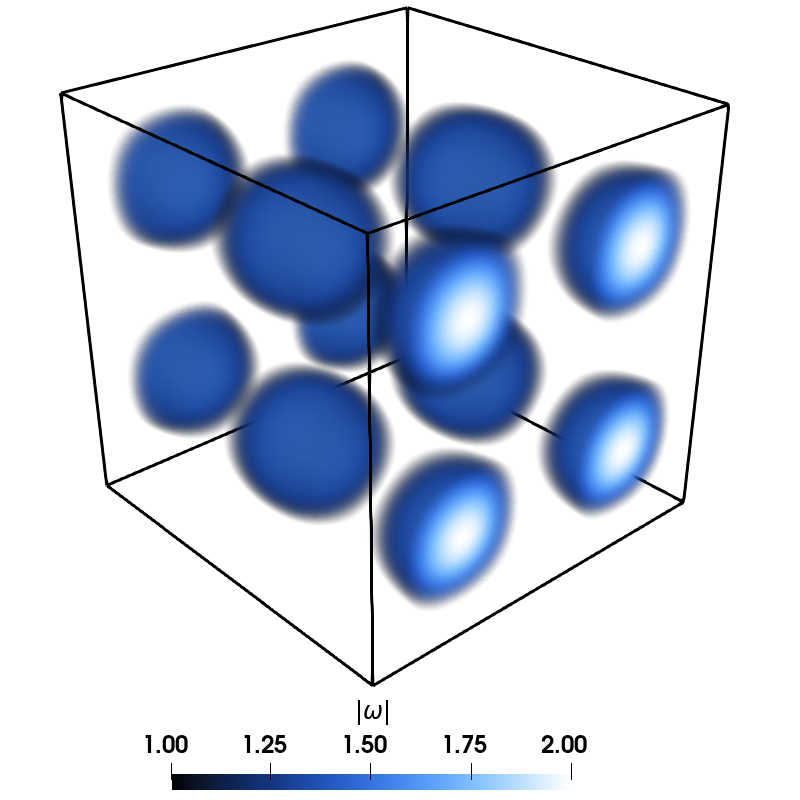}
			\label{vol_rendering0}
		}
		\subfigure[]
		{ 
			\includegraphics[trim = 20mm 0mm 25mm 0mm, clip, width=7cm]{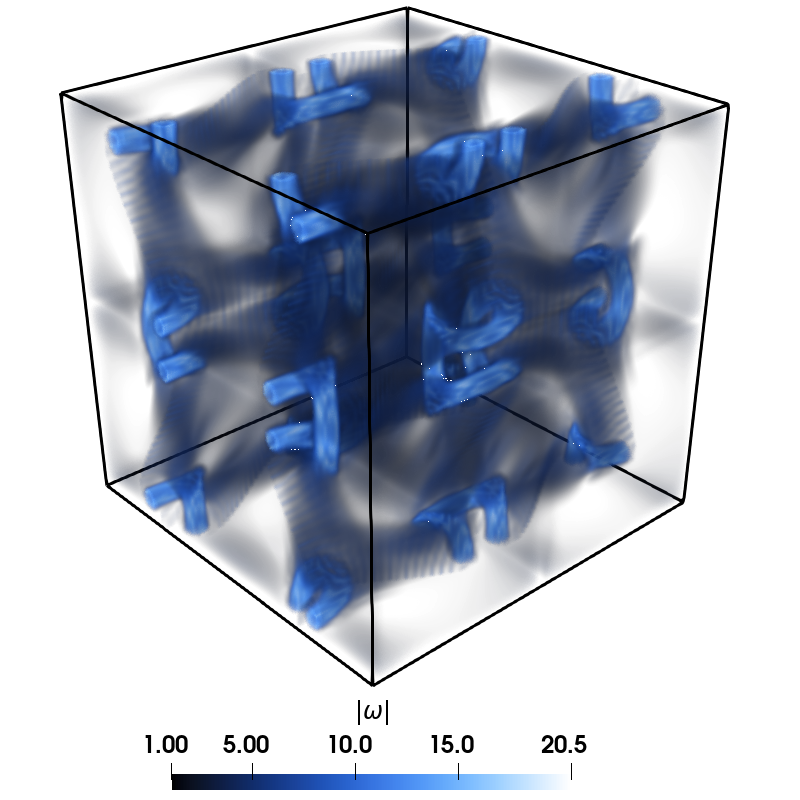}
			\label{vol_rendering5}
		}
		\subfigure[]
		{
			\includegraphics[trim = 20mm 0mm 25mm 0mm, clip, width=7cm]{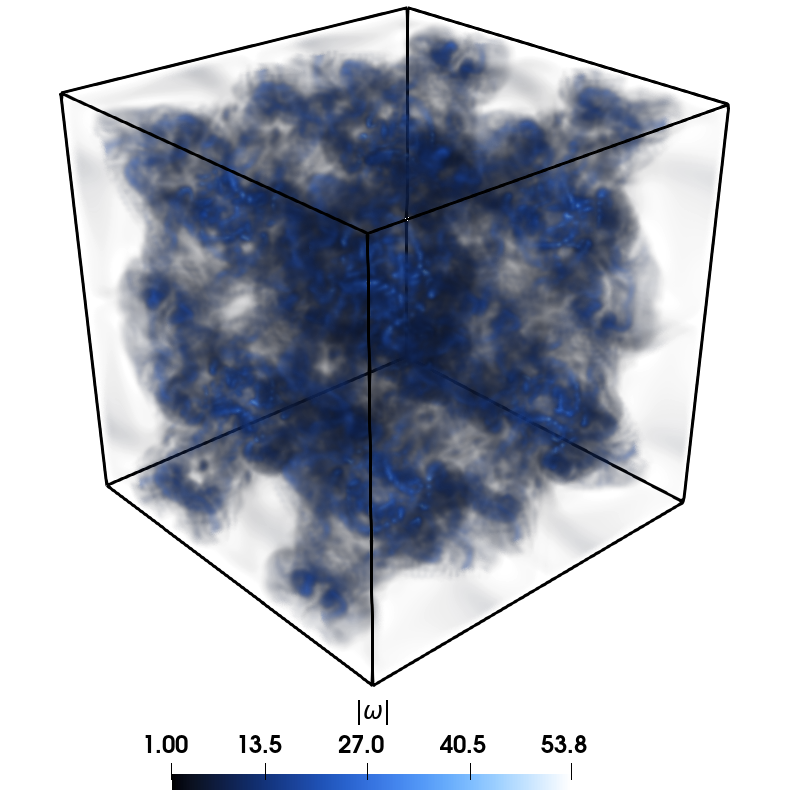}
			\label{vol_rendering10}
		}
		\subfigure[]
		{
			\includegraphics[trim = 20mm 0mm 25mm 0mm, clip, width=7cm]{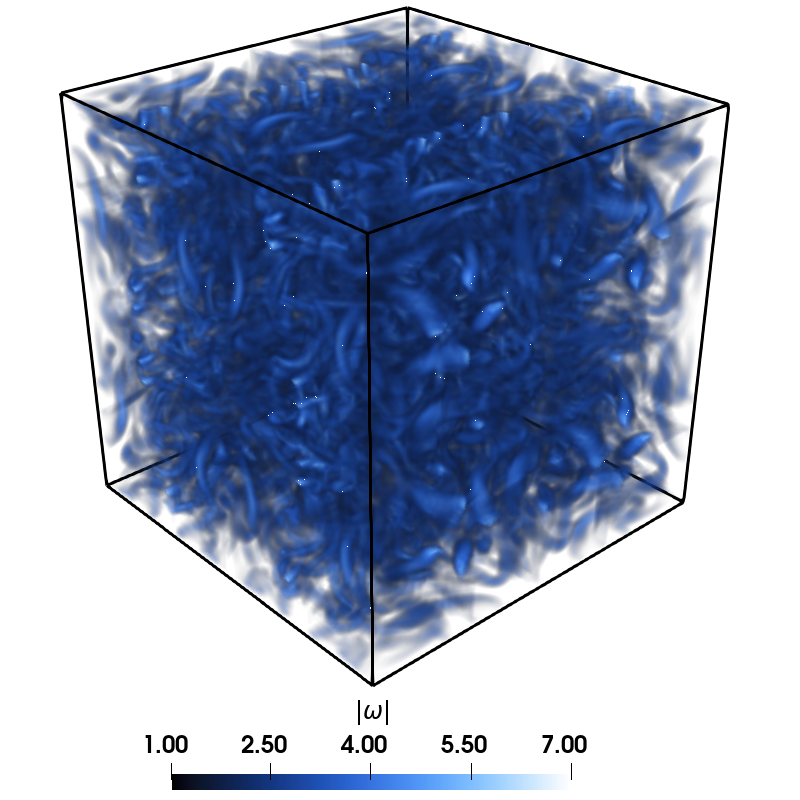}
			\label{vol_rendering20}
		}
		\caption{Volume rendering of the vorticity magnitude, $|\boldsymbol{\omega} |$, at (a)~$t=0$ (b)~$t=5.0$ (c)~$t=9.0$ (d)~$t=20$ for the turbulent Taylor-Green vortex problem from the GPE-based solver with $Ma=0.02$ on the grid of $512^3$.}
		\label{vol_rendering}
	\end{figure}

The second objective of this test case is to investigate the accuracy of the GPE-based weakly compressible approaches to perform scale-resolved simulations of unsteady turbulent flows. The present test case involving the growth and degradation of turbulent structures is one such unsteady state problem. Existing studies~\cite{Dupuy2020,Shi2020} on turbulent flows using GPE focused only on the applicability of statistically-steady configurations like lid-driven cavity, and flows through a channel or a square duct. Some researchers reported results for the present unsteady turbulent flow, but they either employed large eddy simulation~\cite{delorme2017,vermeire2024}, or used insufficiently refined mesh to resolve all the scales~\cite{trojak2022,abdulgafoor2024}. We perform the simulation of this example using $512^3$ mesh that is reported to be sufficient to resolve all the scales~\cite{VanRees2011,bull2015, Abdelsamie2021,carton2014}. Results from our solver are validated against simulations employing high-order schemes reported by Laizet et al.~\cite{laizet2019} and Abdelsamie et al.~\cite{Abdelsamie2021}.

We illustrate the qualitative features of the evolution of flow patterns using the volume rendering of vorticity magnitude~($|\boldsymbol{\omega} |$) at different time instants in figure~\ref{vol_rendering}. Initially, the flow field is observed to be in a laminar regime as in figure~\ref{vol_rendering0}, but it becomes turbulent due to the large value of $Re$. We observe that at around $t=5$ (figure~\ref{vol_rendering5}), the turbulent structures emerge, and a fully turbulent flow pattern is observed at around $t=9$ as shown in figure~\ref{vol_rendering10}. The flow slowly decays thereafter, and we observe these decayed turbulent structures in figure~\ref{vol_rendering20} at $t=20$.

\begin{figure}[H]
		\centering
		\subfigure[]
		{
			\includegraphics[trim = 0mm 0mm 0mm 0mm, clip, width=7.5cm]{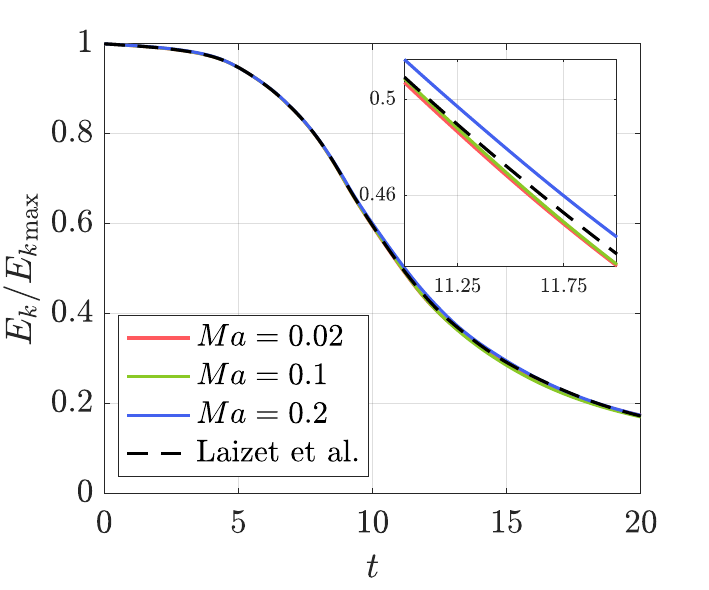}
			\label{tTGV_en_MaStudy}
		}
		\subfigure[]
		{
			\includegraphics[trim = 0mm 0mm 0mm 0mm, clip, width=7.5cm]{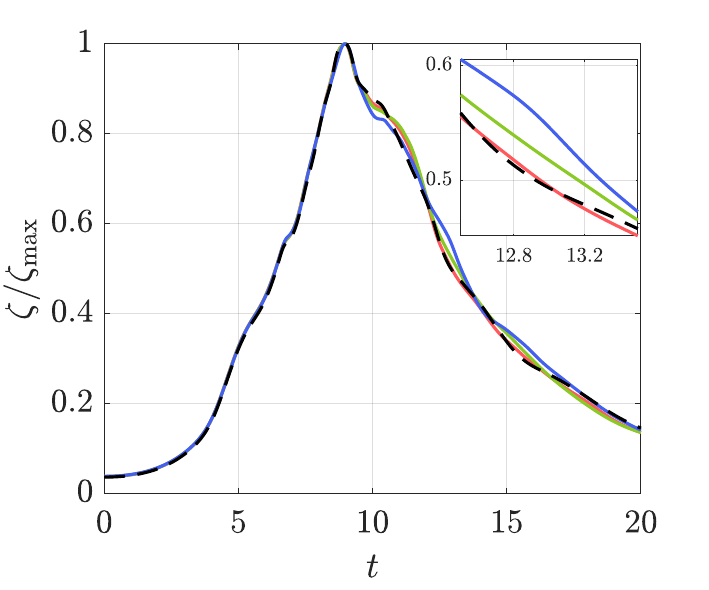}
			\label{tTGV_velDiv_MaStudy}
		}
            \caption{Effect of Mach number on the evolution of (a)~kinetic energy~($E_k$) and (b)~enstrophy~($\zeta$) for the turbulent Taylor-Green vortex problem from the GPE solver using $512^3$ grid.}
		\label{tTGV_ke_512}
	\end{figure}

The quantitative validation of the present solver is carried out by investigating the following monitors, viz., average kinetic energy~($E_k$) and enstrophy~($\zeta$). Both variables are non-dimensionalised by their maximum values and are presented in figure~\ref{tTGV_ke_512}. While results using Ma=0.1 and 0.2 show some deviation, results obtained using $Ma=0.02$ match exceedingly well with the reference values reported by Laizet et al.~\cite{laizet2019}, which is based on a compact sixth-order finite difference scheme. 
    
\begin{figure}[!ht]
		\centering
		\subfigure[]
		{
			\includegraphics[trim = 0mm 0mm 0mm 0mm, clip, width=7.5cm]{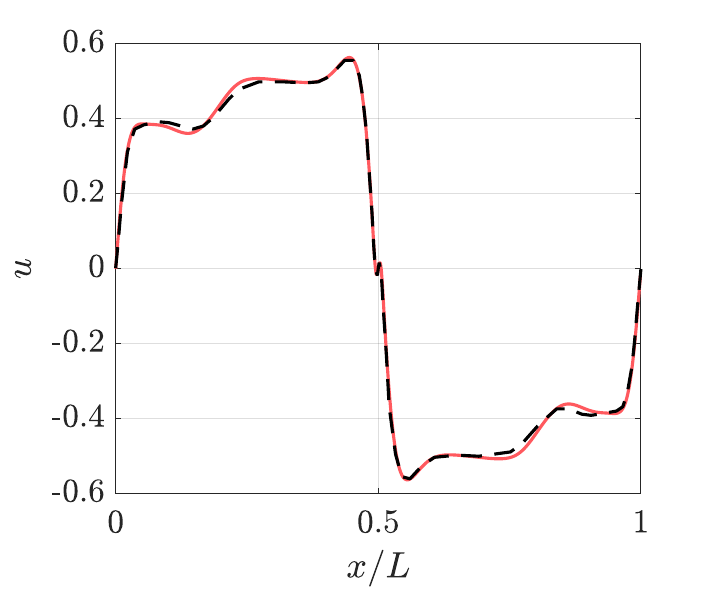}
			\label{tTGV_uVel}
		}
		\subfigure[]
		{
			\includegraphics[trim = 0mm 0mm 0mm 0mm, clip, width=7.5cm]{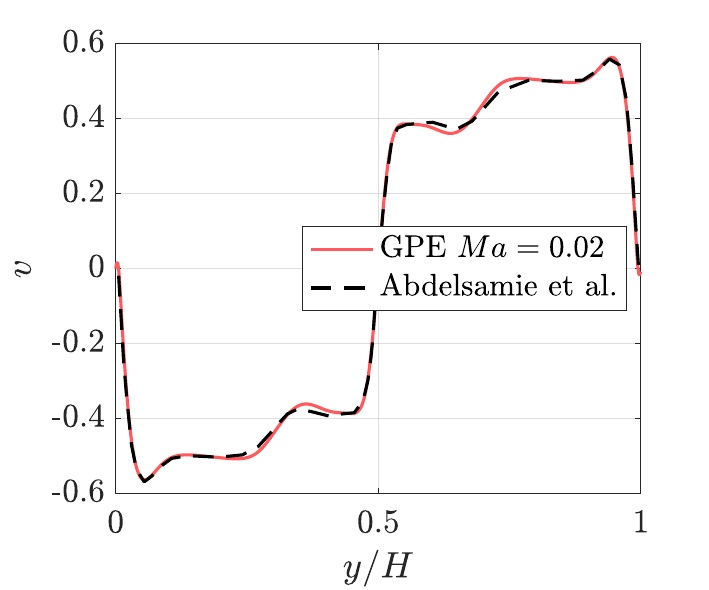}
			\label{tTGV_vVel}
		}
            \caption{Variation of (a)~$x$-component of velocity~($u$) along a line passing through the centre of the domain, parallel to the $x$-axis (distance is non-dimensionalised by length~($L=2\pi$) of the domain) and (b)~$y$-component of velocity~($v$) along a line passing through the centre of the domain, parallel to the $y$-axis (distance is non-dimensionalised by height~($H=2\pi$) of the domain) at $t=12.11$ for the turbulent Taylor-Green vortex problem with a grid of $512^3$ compared with that reported by Abdelsamie et al.~\cite{Abdelsamie2021}.} 
		\label{tTGV_Vel}
	\end{figure}

        In addition to the transient plots, the spatial variation of velocity components along the horizontal and vertical centerline at $t=12.11$ is presented in figure~\ref{tTGV_Vel}. The computed results are compared with the benchmark values of Abdelsamie et al.~\cite{Abdelsamie2021}, which uses a high-order splitting scheme that shows at least seventh-order accuracy. Similar to the previous plots, the results from GPE with the finer mesh of $512^3$ match well with that from the reference.

We underline that despite turbulent flows being highly sensitive to perturbations, the presence of a wide spectrum of length scales, and their complex, unsteady evolution, the mass conservation error doesn't grow with time. Moreover, the spurious dilatation-driven dissipation seems not to intervene with the process of energy cascade, i.e., the generation of small-scale structures by three-dimensional vortex stretching. We can conclude from these results that the GPE-based WCM is capable of accurately simulating unsteady turbulent flows.
\section{Salient points}
\label{salientPoints}

Before concluding, we note some important additional points relevant to WCMs. These are mainly preliminary observations, and further investigations can shed more light on them.
\begin{itemize}
    \item The influence of dilatation-driven dissipation terms is discussed in this paper using a GPE-based WCM. In order to show that this is applicable to EDAC also, we repeated the analysis of the doubly periodic shear layer problem using EDAC. The evolution of total kinetic energy dissipation rate per unit volume at $Ma=0.2$, presented in figure~\ref{fig:EDAC}, shows that the insights provided for GPE are equally relevant for EDAC as well. It would be interesting to see to what extent these spurious dissipation terms affect the family of WCMs~(for example \cite{bigay2017}) that solve the continuity equation for compressible flow instead of a pressure evolution equation as in GPE and EDAC.
\begin{figure}[!ht]
\centering
\includegraphics[trim = 0mm 0mm 0mm 0mm, clip, width=15cm]{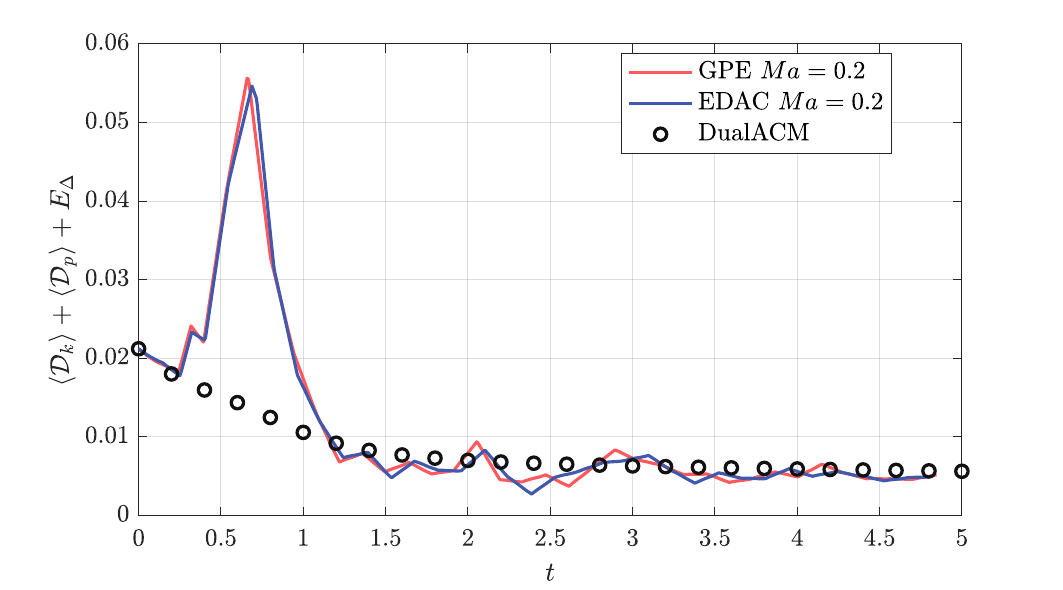}
\caption{Comparison of the total kinetic energy dissipation rate per unit volume for the doubly periodic shear layer problem obtained using GPE and EDAC at $Ma=0.2$.}
\label{fig:EDAC}
\end{figure} 
    \item While the pressure dilatation~($\mathcal{D}_p$) is inherent to WCMs, the kinetic energy dilatation~($\mathcal{D}_k$) arises because of expressing the convective terms of the momentum equation in conservative form. Despite the fact that both forms are equivalent, i.e., $\nabla\cdot(\textbf{u}\otimes\textbf{u})=\textbf{u}\cdot \nabla\textbf{u}$ only for strictly incompressible flows, majority of the studies on GPE~\cite{Toutant2017,Dupuy2020,Shi2020,pan2022,melvin2024,bodhanwalla2024,raghunathan2024} and EDAC~\cite{Clausen2013,delorme2017,kajzer2022,trojak2022,sharma2023,abdulgafoor2024,vermeire2024} used to work with the conservative form. The discussions presented in this paper provide a hint that employing the non-conservative form would be advantageous for WCMs because $\mathcal{D}_k=0$ for this choice. In order to ensure this we performed the DPSL simulation using non-conservative form also. Figure~\ref{fig:convective-compare} compares the kinetic energy and the total dissipation rate evolution of both forms of convective terms, using GPE with $Ma=0.2$. It is clear that the spurious dissipation for the non-conservative form is smaller, consistent with the arguments presented in this paper. Hence, the non-conservative form should be preferred for WCMs.
\begin{figure}[!ht]
    \centering
    \subfigure[]
    {
        \includegraphics[trim = 0mm 0mm 0mm 0mm, clip, width=7.5cm]{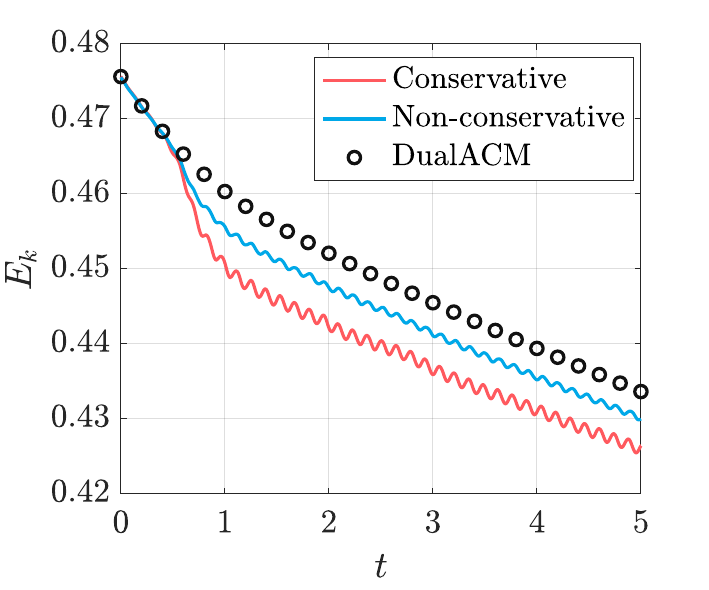}
    }
    \subfigure[]
    {
        \includegraphics[trim = 0mm 0mm 0mm 0mm, clip, width=7.5cm]{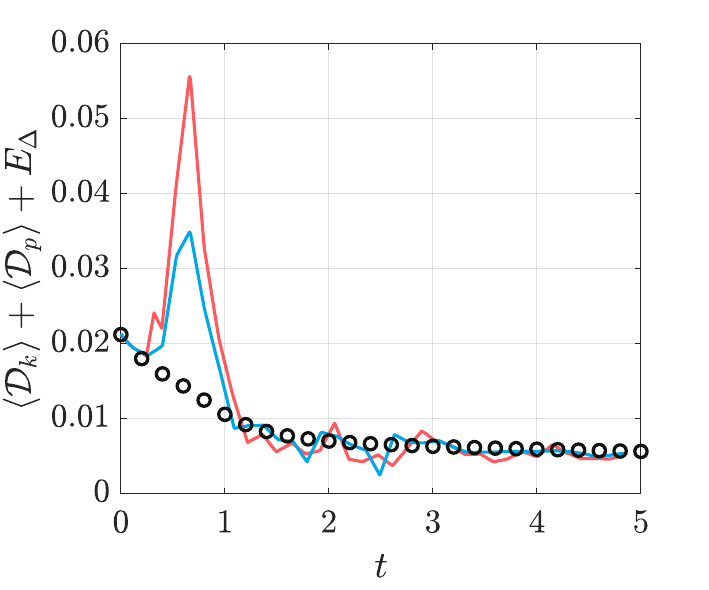}
    }
    \caption{Comparison of conservative and non-conservative forms of the convective equation for the doubly periodic shear layer problem: (a)~kinetic energy, and (b)~total kinetic energy dissipation rate. The results shown are for the GPE solver with $Ma=0.2$.}
    \label{fig:convective-compare}
\end{figure} 
\item The aim of this work is to provide fundamental insights to WCMs. The dilation-driven spurious dissipation terms provided the necessary important information. During this investigation, we found other interesting observations as follows: (i)~the physical process underlying the peaks in $\mathcal{D}_k$ and $\mathcal{D}_p$ that augments the dissipation rate is unclear, and (ii)~the spurious dissipation terms make a negligible contribution to the tTGV problem, despite turbulent flows are sensitive to small perturbations. While exploring the essential physical mechanisms behind these observations will be a curious exercise, it belongs to the fundamental mechanics of weakly compressible fluids and is beyond the scope of the present work.
\item Results presented in this paper have an important practical consequence. For accurately capturing incompressible flows, WCMs mandate $Ma\ll 1$. This enforces a stringent restriction on allowable $\Delta t$, which is a major bottleneck that adversely affects the efficiency of these methods. Until now, WCMs are utilized to solve only canonical problems. The understanding provided in this work offers a hint that a potential means of improving this drawback is to remove the influence of dilatation terms by an appropriate modification of the momentum equation. This could enable relaxing $Ma\ll 1$ condition. Such an advancement can significantly improve the efficiency of WCMs and, together with their capability for parallel computing, can enable solving large-scale industrial problems.
\end{itemize}

\section{Conclusions}
\label{conclusion_section}
The pressure acts as a Lagrange multiplier in incompressible flows to enforce the divergence-free condition on velocity. In weakly compressible methods, aimed at mimicking incompressible flows, the pressure travels at finite speeds that result in non-zero dilatation. This non-adherence to mass conservation corresponding to incompressible flows is a central issue in WCMs. Moreover, it is now well-known that such methods experience enhanced levels of dissipation with increased artificial compressibility. This work aimed to improve the fundamental understanding of WCMs by addressing the effect of mass conservation error on the accuracy of such methods, and unravelling the mechanism behind the dissipation induced by the compressibility. We derived and analysed the kinetic energy equation and identified the dilatation terms that are of crucial importance in governing the behaviour of weakly compressible methods. Results presented in this paper proved that even with the presence of what is considered to be an inadmissibly large mass conservation error, WCMs can accurately capture the flowfield, provided the dilatation-driven spurious dissipation is not significant. Moreover, the dilatation terms can also explain the appearance of non-physical oscillations in time reported for single- and two-phase flow results obtained using WCMs. To the best of our knowledge, this is the first work that provides such key fundamental insights on the behaviour of weakly compressible methods. In addition, we showed that the GPE-based WCM considered in this work is capable of accurately simulating transient incompressible laminar and turbulent flows if the artificial Mach number is small enough.

\section*{Acknowledgments}
    We acknowledge the financial support provided by the DST-SERB Core Research Grant (Project No. CRG/2022/006992).

\bibliographystyle{ieeetr}
\biboptions{sort&compress}
\bibliography{dilatation}

\appendix

\section{Accuracy of DPSL}
\label{app:dpsl}
In this section, we provide additional evidence that the simulation of DPSL using GPE at $Ma=0.02$ provides as accurate results as the reference incompressible flow results generated using DualACM. The comparison between GPE and incompressible flow results for this test case has been presented by Toutant~\cite{Toutant2018}. The additional careful investigation presented here complements this conclusion.

\begin{figure}[!ht]
\centering
    \includegraphics[trim = 0mm 0mm 0mm 0mm, clip, width=15.0cm]{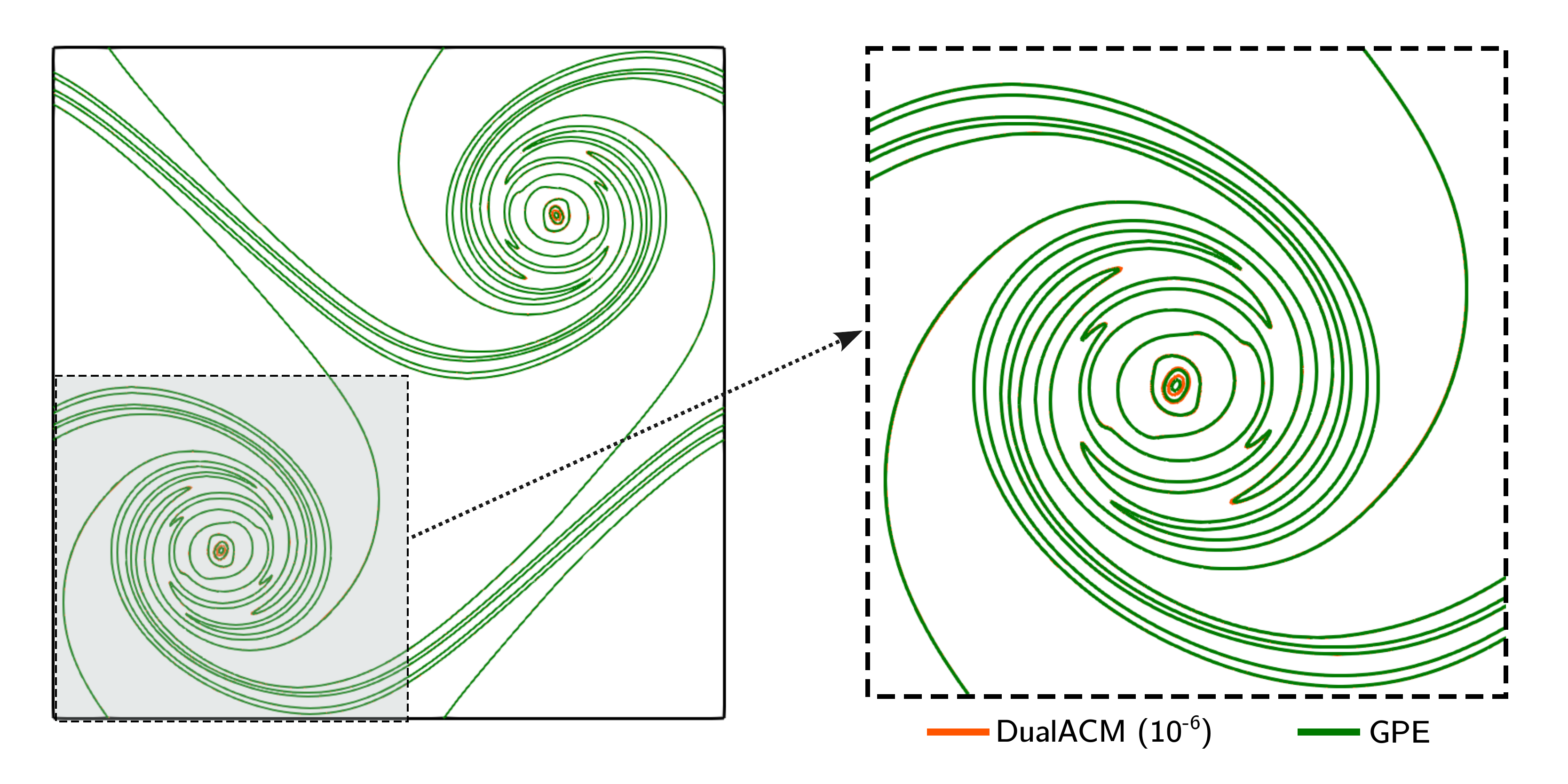}
    \caption{Comparison of vorticity contours from GPE with $Ma=0.02$ and DualACM~($\epsilon=10^{-6}$) for the doubly periodic shear layer test case with a grid of $512 \times 512$ at $t=1$. The vorticity contours are plotted for~$\omega_z = 0, \pm6, \pm18, \pm30, \pm45, \pm55, \pm57, \pm57.5$.}
    \label{dpsl_vort_comp1}
\end{figure}

\begin{figure}[!ht]
    \centering
    \subfigure[]
    {
        \includegraphics[trim = 20mm 0mm 20mm 0mm, clip, width=6cm]{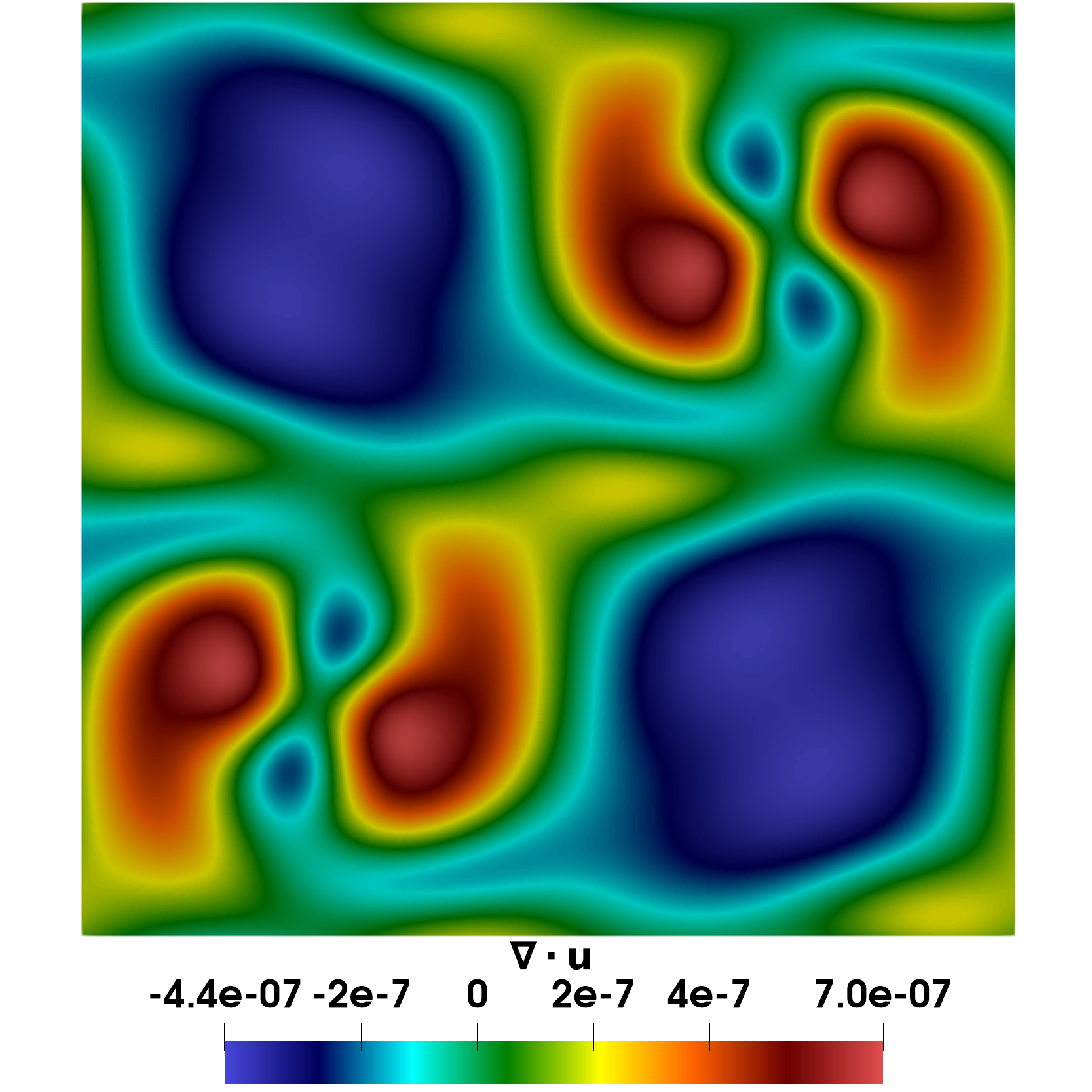}
    }
    \subfigure[]
    {
        \includegraphics[trim = 20mm 0mm 20mm 0mm, clip, width=6cm]{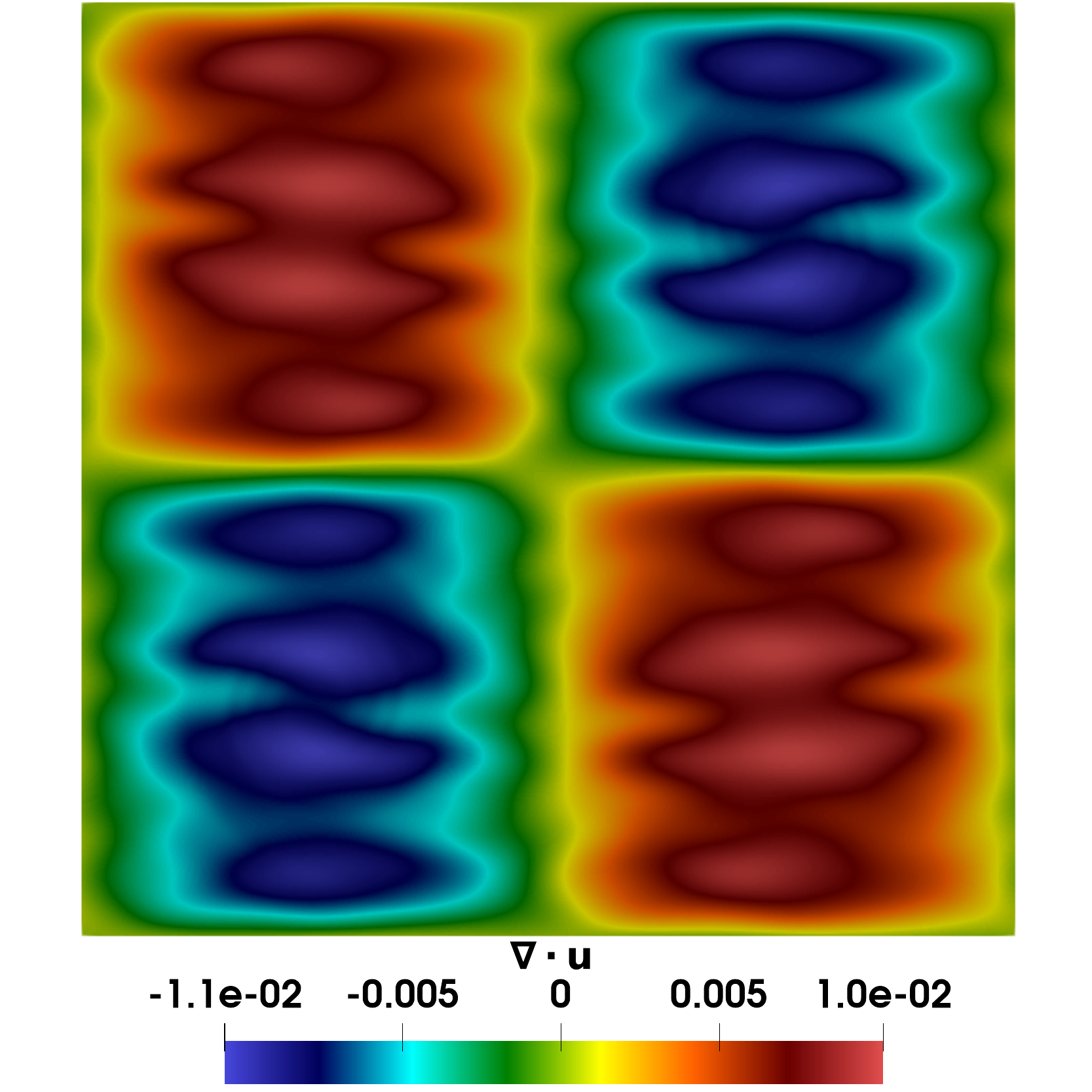}
    }
    \caption{Contours of velocity divergence for the doubly periodic shear layer test case with a grid of 512×512 at $t = 1$ using (a)~DualACM, and (b)~GPE with $Ma=0.02$.}
    \label{dpsl_vel_div}
\end{figure}

The contours of $z-$component of vorticity~($\omega_z$) obtained from the GPE are superimposed on those from the DualACM to facilitate a direct comparison, as presented in figure~\ref{dpsl_vort_comp1}. The contours align so closely that the underlying ones are not visible in most regions. This shows that the results produced by both solvers match exceedingly well, despite the velocity divergence in GPE is approximately four orders of magnitude larger than that of DualACM as illustrated in figure~\ref{dpsl_vel_div}.

        In addition to the qualitative assessment made in figure~\ref{dpsl_vort_comp1}, we also conduct a quantitative comparison between the GPE and DualACM solvers. For this purpose, we compare the maximum value of vorticity and its location from the two solvers along with the reference values reported for a pseudo-spectral method by Hashimoto et al.~\cite{hashimoto2015} as presented in table~\ref{dpsl_maxVort}. It is found that for GPE, the vorticity peaks at 57.57, whereas the corresponding value from  DualACM is 57.72. Evidently, the difference is negligible despite the fact that the velocity divergence differs by four orders of magnitude. Moreover, the difference in the value of maximum vorticity between the GPE and the pseudo spectral method is less than 2\%.
	\begin{table}[!ht]
		\centering
		\caption{Maximum value of vorticity obtained for the doubly periodic shear layer test case for a grid of $512 \times 512$ at $t=1$.}
		\footnotesize
		\begin{tabular}{lllll}
			\: \\
			\hline
			\hline\\
			\textbf{Scheme} & \textbf{Max. value} & \textbf{Location (x,y)}  \\ 
			\hline \\
			GPE  & 57.57  & (0.75, 0.75) \\
			DualACM~($\epsilon=10^{-6}$)  & 57.72  & (0.75, 0.75)  \\
			Pseudo Spectral Method~\cite{hashimoto2015}  & 58.55  & (0.75, 0.75)  \\ \\
			\hline
			\hline
		\end{tabular}
		\label{dpsl_maxVort}
	\end{table}

\end{document}